\documentclass[journal]{IEEEtran}
\usepackage{graphicx}

\usepackage{bm}
\usepackage{amsfonts}
\usepackage{amsthm}
\usepackage{amssymb}
\usepackage{flushend}
\usepackage{color}
\usepackage{mathrsfs}

\usepackage{cite}

\usepackage{booktabs}

\newtheorem{proposition}{Proposition}

\ifCLASSINFOpdf
\else
\fi
\usepackage{amsmath}
\usepackage{eurosym}

\usepackage{amsthm}

\usepackage{algorithm}
\usepackage{algorithmic}
\ifCLASSOPTIONcompsoc
 \usepackage[caption=false,font=normalsize,labelfont=sf,textfont=sf]{subfig}
\else
 \usepackage[caption=false,font=footnotesize]{subfig}
\fi
\begin{document}
%
\title{Joint Radar and Communication Design: Applications, State-of-the-art, and the Road Ahead}
%
%
%
\IEEEspecialpapernotice{(Invited Paper)}
\author{Fan Liu,~\IEEEmembership{Member,~IEEE,}
        Christos Masouros,~\IEEEmembership{Senior~Member,~IEEE,}
        Athina Petropulu,~\IEEEmembership{Fellow,~IEEE,}\\
        Hugh Griffiths,~\IEEEmembership{Fellow,~IEEE}
        and~Lajos Hanzo,~\IEEEmembership{Fellow,~IEEE}
\thanks{This work was supported in part by the Marie Sk{\l}odowska-Curie Individual Fellowship under Grant No. 793345, in part by the Engineering and Physical Sciences Research Council (EPSRC) of the UK Grant number EP/S026622/1, and in part by the UK MOD University Defence Research Collaboration (UDRC) in Signal Processing.}
\thanks{F. Liu, C. Masouros and H. Griffiths are with the Department of Electronic and Electrical Engineering, University College London, London, WC1E 7JE, UK (e-mail: fan.liu@ucl.ac.uk, chris.masouros@ieee.org, h.griffiths@ieee.org).}
\thanks{A. Petropulu is with the Department of Electrical and Computer Engineering, Rutgers, the State University of New Jersey, 94 Brett Road, Piscataway, NJ 08854, United States (e-mail: athinap@rutgers.edu).}
\thanks{L. Hanzo is with the School of Electronics and Computer Science, University of Southampton, Southampton SO17 1BJ, UK. (e-mail: lh@ecs.soton.ac.uk).}
}

\maketitle

\begin{abstract}
For the sake of enhancing the exploitation of the permanently allocated, but potentially under-utilized spectral resources, sharing the frequency bands between radar and communication systems  has attracted substantial attention. More explicitly, there is increasing demand for sharing both the frequency band and the hardware platform between these two functionalities, but naturally, its success critically hinges on high-quality joint sensing and communications. In this paper, we firstly overview the application scenarios and the research progress in the area of communication and radar spectrum sharing, with particular emphasis on: 1) Radar-communication coexistence; 2) Dual-functional radar-communication (DFRC) systems. In the remainder of the paper, we propose a novel transceiver architecture and frame structure for a DFRC base station (BS) operating in the millimeter wave (mmWave) band, using the hybrid analog-digital (HAD) beamforming technique. We assume that the BS is serving a multi-antenna aided user equipment (UE) operating in a mmWave channel, which in the meantime actively detects multiple targets. Note that part of the targets also play the role of scatterers for the communication signal. Given this framework, we then propose a novel scheme for joint target search and communication channel estimation relying on the omni-directional pilot signals generated by the HAD structure. Given a fully-digital communication precoder and a desired radar transmit beampattern, we propose to design the analog and digital precoders under non-convex constant-modulus (CM) and power constraints, such that the BS can formulate narrow beams towards all the targets, while pre-equalizing the impact of the communication channel. Furthermore, we design an HAD receiver that can simultaneously process signals from the UE and echo waves from the targets. By tracking the angular variation of the targets, we show that it is possible to recover the target echoes and mitigate the potential interference imposed on the UE signals by invoking the successive interference cancellation (SIC) technique, even when the radar and communication signals share the equivalent signal-to-noise ratio (SNR). The feasibility and the efficiency of the proposed approaches in realizing DFRC are verified via numerical simulations. Finally, our discussions are summarized by overviewing the open problems in the research field of CRSS.
\end{abstract}


%
\IEEEpeerreviewmaketitle

\section{Introduction}
\subsection{Background}
\IEEEPARstart {G}{iven} the plethora of connected devices and services, frequency spectrum is becoming increasingly congested with the rapid growth of the wireless communication industry. As a consequence, the auction price of the available wireless spectrum has experienced a sharp rise during recent years. For example, since 2015, mobile network operators in the UK have been required to pay a combined annual total of \textsterling 80.3 million for the 900 MHz and \textsterling 119.3 million for the 1800 MHz band, employed for voice and data services using a mix of 2/3/4G technologies\cite{UK_spectrum}. Meanwhile in Germany, the regulator Bundesnetzagentur revealed that the total in the auction of 4 frequency bands for mobile network operators exceeded \euro 5 billion\cite{GM_spectrum}. The US Federal Communications Commission (FCC) completed its first 5G auction, with a sale of 28 GHz spectrum licences raising \$702 million\cite{US_spectrum}. By 2025, the number of connected devices worldwide is predicted to be 75 billion \cite{Connected_devices}, which further emphasizes impending need for extra spectral resources. In view of this, network providers are seeking opportunities to reuse spectrum currently restricted to other applications. The \emph{radar bands} are among at the best candidates to be shared with various communication systems due to the large chunks of spectrum available at radar frequencies \cite{radar_spectrum}.
\\\indent Radar has been developed for decades since its birth in the first half of the 20th century. Modern radar systems are deployed worldwide, with a variety of applications including traffic control, geophysical monitoring, weather observation as well as surveillance for defense and security. Below 10 GHz, a large portion of spectral resources has been primarily allocated to radar, but at the current state-of-the-art new cohabitation options with wireless communication systems, e.g. 5G NR, LTE and Wi-Fi\cite{radar_spectrum}. At the higher frequencies such as the mmWave band, the communication and radar platforms are also expected to achieve harmonic coexistence or even beneficial cooperation in the forthcoming 5G network and beyond. Nevertheless, with the allocation of the available frequency bands to the above wireless technologies, the interference in the radar bands is on the rise, and has raised concerns both from governmental and military organizations for the safeguarding of critical radar operations \cite{FCC_broad,NSF_specees,SSPARC,Ofcom_PSSR,UK_radar_plannning}. To this end, research efforts are well underway to address the issue of communication and radar spectrum sharing (CRSS).
\\\indent In general, there are two main research directions in CRSS: 1) Radar-communication coexistence (RCC) and 2) Dual-functional Radar-Communication (DFRC) system design \cite{7782415}. By considering the coexistence of individual radar and communication systems, the first category of research aims for developing efficient interference management techniques, so that the two systems can operate without unduly interfering with each other. On the other hand, DFRC techniques focus on designing joint systems that can simultaneously perform wireless communication and remote sensing. The joint design benefits both sensing and signalling operations, decongests the RF environment, and allows a single hardware platform for both functionalities. This type of work has been extended to numerous novel applications, including vehicular networks, indoor positioning and secrecy communications \cite{8246850,7060497,5545182}.
\\\indent Below we present existing, or potential application scenarios of CRSS from both civilian and military perspectives.
\begin{table}
\renewcommand{\arraystretch}{1.3}
\caption{List of Acronyms}
\label{table_example}
\centering
\begin{tabular}{@{}|l|r|@{}}
\hline
AoA & Angle of Arrival \\
AoD & Angle of Departure \\
ATC & Air Traffic Control \\
AV & Autonomous Vehicle \\
BS & Base Station \\
CM & Constant Modulus \\
CRSS & Communication and Radar Spectrum Sharing \\
CRB & Cram\'er-Rao Bound \\
CSI & Channel State Information \\
DL & Downlink \\
DP & Downlink Pilot \\
DFRC & Dual-functional Radar-Communication \\
GNSS & Global Navigation Satellite-based Systems \\
GP & Guard Period \\
HAD & Hybrid Analog-Digital Beamforming \\
ICSI & Interference Channel State Information \\
LTE & Long-Term Evolution \\
LPI & Low-probability of Intercept \\
LoS & Line-of-Sight \\
MIMO & Multi-Input-Multi-Output \\
mmWave & Millimeter Wave \\
mMIMO & Massive MIMO \\
MUI & Multi-user Interference \\
MU-MIMO & Multi-user MIMO \\
NR & New Radio \\
NSP & Null-space Projection \\
NLoS & Non Line-of-Sight \\
PRF & Pulse Repetition Frequency \\
PRI & Pulse Repetition Interval \\
RCC & Radar-Communication Coexistence \\
RCS & Radar Cross Section \\
RF & Radio Frequency \\
RFID & Radio Frequency Identification \\
SIC & Successive Interference Cancellation \\
SINR & Signal-to-Interference-plus-Noise Ratio \\
SNR & Signal-to-Noise Ratio \\
SVD & Singular Value Decomposition \\
TDD & Time-division Duplex \\
UAV & Unmanned Areial Vehicle \\
UE & User Equippment \\
UL & Uplink \\
UP & Uplink Pilot \\
V2X & Vehicle-to-Everything \\
WPS & WiFi Positioning System \\
\hline
\end{tabular}
\end{table}
\subsection{Civilian Applications}
\emph{1) Coexistence of individual radar and wireless systems}
\\\indent As discussed above, CRSS has originally been motivated by the need for the coexistence of radar and commercial wireless systems. Next, we provide examples of coexisting systems in various bands.
\begin{itemize}
    \item \emph{L-band (1-2 GHz)}: This band is primarily used for long-range air-surveillance radars, such as Air Traffic Control (ATC) radar, which transmits high-power pulses with broad bandwidth. The same band, however, is also used by 5G NR and FDD-LTE cellular systems as well as the Global Navigation Satellite System (GNSS) both in their downlink (DL) and uplink (UL) \cite{rad_lte}.
    \item \emph{S-band (2-4 GHz)}: This band is typically used for airborne early warning radars at considerably higher transmit power\cite{7462190}. Some long-range weather radars also operate in this band due to moderate weather effects in heavy precipitation \cite{radar_spectrum}. Communication systems present in this band include 802.11b/g/n/ax/y WLAN networks, 3.5 GHz TDD-LTE and 5G NR \cite{rad_wifi}.
    \item \emph{C-band (4-8 GHz)}: This band is very sensitive to weather patterns. Therefore, it is assigned to most types of weather radars for locating light/medium rain \cite{radar_spectrum}. On the same band operate radars used for battlefield/ground surveillance and vessel traffic service (VTS)\cite{radar_spectrum}. Wireless systems in this band mainly include WLAN networks, such as 802.11a/h/j/n/p/ac/ax \cite{wiki_wlan}.
    \item \emph{MmWave band (30-300 GHz)}\footnote{Typically, communication systems operated close to 30GHz (e.g. 28GHz) are also referred as mmWave systems.}: This band is conventionally used by automotive radars for collision detection and avoidance, as well as by high-resolution imaging radars \cite{7786130}. However, it is bound to become busier, as there is a huge interest raised by the wireless community concerning mmWave communications, which are soon to be finalized as part of the 5G NR standard \cite{6736750}. Currently, the mmWave band is also exploited by the 802.11ad/ay WLAN protocols \cite{wiki_wlan}.
\end{itemize}
Among the above coexistence cases, the most urgent issues arise due to interference between base stations and ATC radars \cite{rad_lte}. Early in 2012, a report pointed out that airport radar could delay the deployment of LTE in Southeast England, especially at the main London gateways \cite{radar_lte_delay}. In the forthcoming 5G network, the same problem still remains to be resolved. For reasons of clarity, we summarize the above coexistence cases in TABLE II.
\begin{table*}
\centering
\caption{Radar-communication coexistence cases}
\begin{tabular}{@{}lll@{}}
\toprule
\textbf{Frequency Band} & \textbf{Radar Systems} & \textbf{Communication Systems} \\ \midrule
\textbf{L-band (1-2GHz)} & Long-range surveillance radar, ATC radar & LTE, 5G NR \\ \midrule
\textbf{S-band (2-4GHz)} & \begin{tabular}[c]{@{}l@{}}Moderate-range surveillance radar, ATC radar, \\ airborne early warning radar\end{tabular} & \begin{tabular}[c]{@{}l@{}}IEEE 802.11b/g/n/ax/y WLAN, \\ LTE, 5G NR\end{tabular} \\ \midrule
\textbf{C-band (4-8GHz)} & \begin{tabular}[c]{@{}l@{}}Weather radar, ground surveillance radar, \\ vessel traffic service radar\end{tabular} & IEEE 802.11a/h/j/n/p/ac/ax WLAN \\ \midrule
\textbf{MmWave band (30-300GHz)} & Automotive radar, high-resolution imaging radar & IEEE 802.11ad/ay WLAN, 5G NR \\ \bottomrule
\end{tabular}
\end{table*}
\\\indent \emph{2) 5G mmWave localization for vehicular networks}
\\\indent In next-generation autonomous vehicle (AV) networks, vehicle-to-everything (V2X) communication will require low-latency Gbps data rates; while general communications can deal with hundreds of ms delays, AV-controlled critical applications require delays of the order of tens of ms \cite{8246850}. In the same scenario, radar sensing should be able to provide robust, high-resolution obstacle detection on the order of a centimeter. At the time of writing, vehicular localization and networking schemes are mostly built upon GNSS or default standards such as dedicated short-range communication (DSRC)\cite{5888501} and the D2D mode of LTE-A \cite{7786130}. While these approaches do readily provide basic V2X functionalities, they are unable to fulfill the demanding requirements mentioned above. As an example, the 4G cellular system provides the localization information at an accuracy on the order of 10m, at a latency often in excess of 1s, and is thus far from ensuring driving safety \cite{8246850}.
\\\indent It is envisioned that the forthcoming 5G technology, exploiting both massive MIMO antenna arrays and the mmWave spectrum, will be able to address the future AV network requirements \cite{6515173,7400949}. The large bandwidth available in the mmWave band would not only enable higher data rates, but would also significantly improve range resolution. Furthermore, large-scale antenna arrays are capable of formulating ``thumbtack-like" beams that accurately point to the directions of interest; this could compensate for the path-loss encountered by mmWave signals, while potentially enhancing the angle of arrival (AoA) estimation accuracy. More importantly, as the mmWave channel is characterized by having only a few multipath components, there is far less clutter interference imposed on target echoes than that of the rich scattering channel encountered in the sub-6GHz band, which is thus beneficial for localization of vehicles \cite{8246850}.
\\\indent For all advantages mentioned in Sec. I-A, it would make sense to equip vehicle or road infrastructure sensors with joint radar and communication functionalities. While the current DFRC system has considered sensors with dual functionality, those were mainly for the lower frequency bands, and cannot be easily extended to the V2X scenario. However, several problems need to be investigated in that context, such as specific mmWave channel models and constraints.
\\\indent \emph{3) Wi-Fi based indoor localization and activity recognition}
\\\indent Indoor positioning technologies represent a rapidly growing market, and thus are attracting significant research interest \cite{7060497,7303962}. While the GNSS is eminently suitable for outdoor localizations, its performance degrades drastically in an indoor environment. To address the above issue, Wi-Fi based positioning system (WPS) constitutes promising solutions, as a benefit of their low cost and ubiquitous deployment, while requiring no additional hardware \cite{7060497}. Essentially, WPS can be viewed as a type of \emph{passive radar}, which locates the target based on the received signals sent by the user equipment (UE). In general, the UE is localized based on the estimation of its time of arrival (ToA) and AoA parameters. Alternatively, the localization information can also be obtained by measuring the received signal strength (RSS) and by exploiting its fingerprint properties (frequency response, signal strength regarding the I/Q channel, etc.), which are then associated with a possible location in a pre-measured fingerprint database \cite{6042868,6244790}.
\\\indent To gain more detailed information concerning the target such as the human behavior, the receiver can process the signal reflected/scattered by the human body, based on specific transmitted signals. This system is more similar to a \emph{bistatic radar} than to conventional WPS. The micro-Doppler shift caused by human activities can be further extracted from the channel state information (CSI) of the Wi-Fi, and analyzed for recognizing human actions \cite{8360863,7944276}. Potential applications of such techniques go far beyond the conventional indoor localization scenarios, which include health-care for elderly people, contextual awareness, anti-terrorism actions and Internet-of-Things (IoT) for smart homes \cite{7426551,8360863,7046290}. It is worth highlighting that a similar idea has been recently applied by the Soli project as part of the Google Advanced Technology and Projects (ATAP), where a mmWave radar chip is sophistically designed for finger-gesture recognition by exploiting the micro-Doppler signatures, hence enabling touchless human-machine interaction \cite{google_soli}.
\\\indent The above technology can be viewed as a particular radar/sensing functionality incorporated into a Wi-Fi communication system, which again falls into the area of DFRC. Consequently, sophisticated joint signal processing approaches need to be developed for realizing simultaneous localization and communications.
\\\indent \emph{4) Unmanned aerial vehicle (UAV) communication and sensing}
\\\indent UAVs have been proposed as aerial base stations to a range of data-demanding scenarios such as concerts, football games, disasters and emergency scenarios\cite{8531711}. It is worth noting that in all of these applications, communication and sensing are a pair of essential functionalities. In contrast to the commonly-used camera sensor on the typical UAV platforms which are sensitive to environmental conditions, such as light intensity and weather, radio sensing is more robust and could thus be incorporated into all-weather services. Additionally, radio sensing could be adopted in drone clusters for formation flight and collision avoidance \cite{1428700}. While both communication and sensing techniques have been individually investigated over the past few years, the dual-functional design aspect remains widely unexplored for UAVs. By the shared exploitation of the hardware between sensors and transceivers, the payload on the UAV is minimized, which increases its mobility/flexibility, while reducing the power consumption \cite{7470933}.
\\\indent \emph{5) Others}
\\\indent Apart from the aforementioned research contributions, there are also a number of interesting scenarios, where CRSS based techniques could find employment, which include but are not limited to:
\begin{itemize}
\item \emph{Radio Frequency Identification (RFID):} A typical RFID system consists of a reader, reader antenna array and tags. Tags can either be passive or active depending on whether they carry batteries. To perform the identification, the reader firstly transmits an interrogation signal to the tag, which is modulated by the tag and then reflected back to the reader, giving a unique signature generated by the particular variation of the tag's antenna load \cite{6646211}. The RFID based sensing is carried out by establishing a cooperative communication link between the reader and the tag. Hence this combines radar and communication techniques to a certain degree.
\item \emph{Medical sensors:} To monitor the health conditions of patients, bio-sensors may be embedded in the human body. As these sensors support only low-power sensing relying on their very limited computational capability, the measured raw data has to be transmitted to an external device for further processing. The most reliable solution for joint sensing and communication is yet to be explored in that scenario \cite{bio_sen}. This, however, requires more interdisciplinary approaches.
\item \emph{Radar as a relay:} In contrast to classic wireless communications, most of radar waveforms are high-powered and strongly directional. These properties make the radar a suitable communication relay, which can amplify and forward weak communication signals to remote users \cite{6875553}. Again, joint radar and communication relaying can play a significant role here.
\end{itemize}
\subsection{Military Applications}
\emph{1) Multi-function RF systems}
\\\indent The development of shipborne and airborne RF systems, including communication, electronic warfare (EW) and radar, has historically been isolated from each other. The independent growth of these sub-systems led to significant increase in the volume and weight of the combat platform, as well as in the size of the antenna array. This results in a larger radar cross-section (RCS) and a consequently increased detectability by adversaries. Moreover, the addition of such sub-systems will inevitably cause electromagnetic compatibility issues, which may impose serious mutual interference on the existing subsystems. To address these problems, the Advanced Multi-function Radio Frequency Concept (AMRFC) project was launched by the Defense Advanced Research Projects Agency (DARPA) in 2005, whose aim was to design integrated RF systems capable of simultaneously supporting multiple functions mentioned above\cite{858893,1406306}. In 2009, the Office of Naval Research (ONR) sponsored a follow-up project namely the Integrated Topside (InTop) program \cite{6127573}, with one of its goals to further develop wideband RF components and antenna arrays for multi-function RF systems based on the outcome of AMRFC.
\\\indent Clearly, the fusion of the radar and the communication subsystems is at the core of the above research. By realizing this, a dedicated project named as ``Shared Spectrum Access for Radar and Communications (SSPARC)'' was funded by the DARPA in 2013, and was further proceeded into the second phase in 2015 \cite{SSPARC}. The purpose of this project is to release part of the sub-6GHz spectrum which is currently allocated to radar systems for shared use by radar and wireless communications. By doing so, SSPARC aims for sharing the radar spectrum not only with military communications, but also with civilian wireless systems, which is closely related to the coexistence cases discussed in Sec. I-B.
\\\indent \emph{2) Military UAV applications}
\\\indent In addition to the civilian aspect mentioned above, UAVs have also been considered as an attractive solution to a variety of military missions that require high mobility, flexibility and covertness. Such tasks include search and rescue, surveillance and reconnaissance as well as electronic countermeasures \cite{1677946,6105461,7935430}, all of which need both sensing and communication operations. Similar to its civilian counterpart, the integration of the two functionalities could significantly reduce the payload as well as the RCS of the UAV platform.
\\\indent On the other hand, UAVs can also be a threat to both infrastructures and people, as it might be used to carry out both physical and cyber attacks. Moreover, even civilian UAVs can impose unintentional but serious danger if they fly into restricted areas \cite{8490190}. To detect and track unauthorized UAVs, various techniques such as radar, camera and acoustic sensors have been employed. Nevertheless, a dedicated equipment specifically conceived for sensing UAVs could be expensive to deploy \cite{5592987}. Therefore, there is a growing demand to utilize existing communication systems, such as cellular BSs, to monitor unauthorized UAVs while offering wireless services to authorized UEs, which needs no substantial extra hardware and thus reduces the cost \cite{8624565}. By modifying BSs for acting as low-power radars, the future Ultra Dense Network (UDN) having a large number of cooperative micro BSs can be exploited as the urban air defense system, which provides early warning of the incoming threats.
\begin{table}
\caption{Applications of the CRSS technology}
\centering
\label{my-label1}
\begin{tabular}{@{}ll@{}}
\toprule
\textbf{Civilian Applications} & \begin{tabular}[c]{@{}l@{}}Radar-comms coexistence, V2X network, \\ WiFi localization, UAV comms and sensing, \\ RFID, Medical sensors,  Radar relay, etc.\end{tabular} \\ \midrule
\textbf{Military Applications} & \begin{tabular}[c]{@{}l@{}}Multi-function RF system, LPI comms, \\ UAV comms and sensing, Passive radar, etc.\end{tabular}                     \\ \bottomrule
\end{tabular}
\end{table}
\\\indent \emph{3) Radar-assisted low-probability-of-intercept (LPI) communication}
\\\indent The need for covert/secrecy communication has emerged in many defense-related applications, where sensitive information such as the locations of critical facilities should be protected during the transmission. The probability of intercept is thus defined as a key performance metric for secrecy communications. Conventionally, LPI is achieved by frequency/time hopping or spread-spectrum methods, which require vast time and frequency resources \cite{1146255,1146256}. From a CRSS viewpoint, however, a more cost-efficient approach would be to embed the communication signal into radar echo waves to mask the data transmission \cite{5545182,6081358,7376230}.
\\\indent A general model for the above scenario is composed by a RF tag/transponder within a collection of scattered targets and a radar transceiver. To elaborate briefly, the radar firstly emits a probing waveform, which is captured by the RF tag on its way to the targets. The tag then remodulates the radar signal with communication information and sends it back to the radar, which is naturally embedded in the reflected radar returns \cite{5545182}. The communication waveform should be appropriately designed by controlling its transmit power and the correlation/similarity with the radar waveform. As such, the communication signal can be hard to recognize at the adversary's side, since it hides itself behind the random clutters and echoes. Nevertheless, it can be easily decoded at the radar by exploiting some {\emph{a priori}} knowledge \cite{6081358}. Accordingly, a number of performance trade-offs among radar sensing, communication rate and information confidentiality can be achieved by well-designed waveforms and advanced signal processing techniques.
\\\indent \emph{4) Passive radar}
\\\indent From a broader viewpoint, passive radar, which exploits scattered signals gleaned from non-cooperative communication systems, could be classified as a special type of CRSS technology. Such illumination sources can be television signals, cellular BSs and digital video/audio broadcasting (DVB/DAB) \cite{7944264}. To detect a target, the passive radar firstly receives a reference signal transmitted from a direct LoS path (usually referred as ``reference channel'') from the above external TXs. In the meantime, it listens to the scattered counterpart of the same reference signal that is reflected by potential targets (referred as ``surveillance channel'') \cite{passive_radar_book}. Note that these scattered signals contain target information similarly to the case of active radars. As a consequence, the related target parameters can be estimated by computing the correlation between signals gleaned from the two channels.
\\\indent The passive radar is known to be difficult to locate or be interfered, since it remains silent when detecting targets, and hence it is advantageous for covert operations. Furthermore, it requires no extra time/frequency resources, leading to a cost that is significantly lower than that of its conventional active counterparts. For this reason, it has been termed ``green radar" \cite{passive_radar_book}. Nonetheless, it may suffer from poor reliability due to the facts that the signal used is not specifically tailored for target detection, and that the transmit source is typically not under the control of the passive radar \cite{passive_radar_book}. To further improve the detection probability while guaranteeing a satisfactory communication performance, joint waveform designs and resource allocation approaches could be developed by invoking CRSS techniques \cite{7962141}.
\\\indent For clarity, we summarize the aforementioned application scenarios of CRSS technologies in TABLE III.

\section{Literature Review}
In this section, we review the recent research progress in the area of CRSS. We will firstly introduce the coexistence approaches for individual radar and communication systems, and then the family of the dual-functional radar-communication system designs.
\subsection{Radar-Communication Coexistence (RCC)}
\emph{1) Opportunistic spectrum access}
\\\indent From the perspective of cognitive radio, a straightforward approach is the so-called opportunistic spectrum access, in which the radar is regarded as the primary user (PU) of the spectrum, whereas the communication system plays the role of the secondary user (SU). Such methods typically require the SU to sense the spectrum, by which a transmission opportunity is obtained when the spectrum is unoccupied. To avoid imposing interference on the radar, the communication system has to control its power to ensure that the radar's interference-to-noise ratio (INR) does not become excessive \cite{4557030}. A similar approach has been adopted in \cite{6331681} for the coexistence of a rotating radar and a cellular BS. In this scenario, the mainlobe of the radar antenna array rotates periodically to search for potential targets. The BS is thus allowed to transmit only when it is in the sidelobe of the radar. Under this framework, the minimum distance between the two systems is determined given the tolerable INR level, and the communication performance is also analyzed in terms of the DL data rate.
\\\indent Although being easily implemented in realistic scenarios, the above approaches are unable to fully exploit the shared use of the spectrum. This is because the communication system can only operate under certain circumstances, namely when the radar is not occupying the frequency and the spatial resources. Additionally, the above contributions do not easily extend to facilitate coexistence with MIMO radar. Unlike conventional radars, the MIMO radar transmits omnidirectional waveforms to search for unknown targets across the whole space, and formulates directional beams to track known targets of interest \cite{4350230,li2008mimo}. Consequently, it is hard for the BS to identify the sidelobes of the MIMO radar, since the radar beampattern may change randomly along with the movement of the targets. Therefore, more powerful techniques such as transmit precoding design are required to cancel the mutual interference.
\\\indent \emph{2) Interference channel estimation}
\\\indent Before designing a transmit precoder, the interference channel state information (ICSI), i.e. the information on the channel where the mutual interference signals propagate, should be firstly obtained. Conventionally, this information is obtained by exploiting the received pilot signals received from the BS at the radar, which might consume extra computational and signaling resources \cite{7814210}. As another option, the authors of \cite{7953658} proposed to build a dedicated control center connected to both systems via wireless or backhaul links, which would carry out all the coordinations including ICSI estimation and transmit precoding design. In cases where the radar has priority, such a control center would be part of the radar \cite{7814210}. However, such a method would involve significant overhead. A novel channel estimation approach has been proposed in \cite{ICSI_est} by exploiting the radar probing waveform as the pilot signal, where the radar is oblivious to the operation of the communication system. Since the radar randomly changes its operational mode from searching to tracking, the BS has to firstly identify the working modes of the radar by hypothesis testing methods, and then estimate the channel.
\\\indent \emph{3) Closed-form precoder design}
\\\indent After estimating the interference channel, the precoder can be designed at either the radar or the communication's side. Similar to the zero-forcing (ZF) precoding of classic MIMO communication, a simple idea is the so-called null-space projection (NSP) \cite{7089157}, which typically requires the radar to have the knowledge of the ICSI. In the NSP scheme, the radar firstly obtains the right singular vectors of the interference channel matrix by singular value decomposition (SVD), and then constructs an NSP precoder relying on those vectors associated with the null space of the channel. The precoded radar signal is projected onto the null-space of the channel, so that the interference power received at the BS is strictly zero. However, such a precoder might lead to serious performance losses of the MIMO radar, for example by eroding the spatial orthogonality of the searching waveform. To cope with this issue, the authors of \cite{6831613} designed a carefully adjusted threshold for the singular values of the channel matrix and then formulated a relaxed NSP precoder by the right singular vectors associated with singular values that are smaller than the threshold. By doing so, the radar performance can be improved at the cost of increasing the interference power received at the BS.
\\\indent Despite the above-mentioned benefits, there are still a number of drawbacks in NSP based approaches. For instance, the interference power can not be exactly controlled, since it is proportional to the singular values of the random channel. Additionally, since the target's response might fall into the row space of the communication channel matrix, it will be zero-forced by the NSP precoder and as a consequence, be missed by the radar. Fortunately, these disadvantages could be overcome by use of convex optimization techniques, which optimize the performance of both systems under controllable constraints.
\\\indent \emph{4) Optimization based designs}
\\\indent Pioneering effort on optimization based beamforming/signaling for the RCC is the work in \cite{7470514}, where the coexistence of a point-to-point (P2P) MIMO communication system and a Matrix-Completion MIMO (MC-MIMO) radar is considered. As a computationally efficient modification of the MIMO radar, the MC-MIMO radar typically employs a sub-sampling matrix to sample the receive signal matrix of the target echoes, and approximately recovers the target information using the matrix completion algorithm \cite{7470514}. The random sub-sampling at the radar receive antennas modulates the interference channel, and increases its null space. This gives the opportunity to the communication system to design its precoding scheme so that it minimizes the interference caused to the radar. In \cite{7470514}, the covariance matrix of the communication signal and the sub-sampling matrix of the MC-MIMO radar are jointly optimized, subject to power and capacity constraints. The corresponding optimization problem is solved via Lagrangian dual decomposition and alternating minimization methods. By taking realistic constraints into consideration, the authors further introduce signal-dependent clutter into the coexistence scenario in \cite{7953658}, which has to be reduced to maximize the effective SINR of the radar while guaranteeing the communication performance. It has been also pointed out in \cite{7953658} that while the interference imposed by the communication system onto the radar is persistent, the interference inflicted by the radar upon the communication link is intermittent. By realizing this, the authors of \cite{8048004} have considered the coexistence issues of a communication system and a pulsed radar, and quantified the communication rate as the weighted sum of the rates with and without the radar interference, which is named as the {\emph{compound rate}}. The authors then formulate an optimization problem to maximize the rate subject to power and radar SINR constraints. It is worth noting that this problem can be solved in closed-form when the radar interference satisfies certain conditions.
\\\indent To address the coexistence problem of the MIMO radar and the multi-user MIMO (MU-MIMO) communication system, the authors of \cite{7898445} have proposed a robust beamforming design at the MIMO BS when the ICSI between the radar and the communication system is imperfectly known. An optimization problem is formulated for maximizing the detection probability of the radar, while guaranteeing the power budget of the BS and the SINR of the DL users. Cui et al. \cite{8445973} have further proposed an interference alignment based transmit precoding design with special emphasis on the degree of freedom (DoF), under the scenario where multiple communication users coexist with multiple radar users. More recently, a constructive interference based beamforming design has been proposed for the coexistence scenario \cite{8355705}, where the known DL multi-user interference (MUI) is utilized for enhancing the useful signal power. As a result, the SINR of the DL users is significantly improved compared to that of \cite{7898445} given the same transmit power budget. We refer readers to \cite{7103338} for more details on the topic of interference exploitation.
\\\indent \emph{5) Receiver designs}
\\\indent We end this section by briefly reviewing the receiver designs conceived for the coexistence of radar and communications. The aim of such a receiver is to estimate the target parameters in the presence of the communication interference, or to demodulate the communication data while cancelling the radar interference, depending on which side it belongs to. To the best of our knowledge, most of the existing research is focused on the second type, i.e., on the design of receivers for communication systems.
\\\indent In \cite{8233171}, the authors consider a spectrum sharing scenario in which a communication receiver coexists with a set of radar/sensing systems. In contrast to the cooperative scenarios discussed in the relevant literature \cite{7089157,7470514,7953658}, the authors of \cite{8233171} assume that the only information available at the communication system is that the interfering waveforms impinging from the radars fall into the subspace of a known dictionary. Given the sparse properties of both the radar interference and the communication demodulation errors, several optimization algorithms have been conceived for simultaneously estimating the radar interference, whilst demodulating the communication symbols based on compressed sensing (CS) techniques. It is shown that the associated optimization problems can be efficiently solved via non-convex factorization and conjugate gradient methods.
\\\indent In a typical coexistence scenario, the communication system periodically receives radar interfering pulses having high amplitudes and short durations, which implies that a narrow-band communication receiver experiences radar interference as an approximately constant-amplitude additive signal. Due to the slow variation of the radar parameters, this amplitude can be accurately estimated. Nevertheless, the phase shift of the interfering signal is sensitive to the propagation delay, thus is difficult to obtain. In \cite{8332962}, the authors exploit the assumption that the amplitude of the radar interference is known to the communication receiver, whereas the phase shift is unknown and uniformly distributed on $\left[0,2\pi\right]$. With the presence of the interfering signal receiving from the radar, a pair of communication-related issues have been studied. The first one is how to formulate the optimal decision region on a given constellation based on the maximum likelihood (ML) criterion. The second one, on the other hand, is how to design self-adaptive constellations that optimize certain metrics, namely the communication rate and the symbol error rate (SER). It is observed via numerical simulations that the optimal constellation tends to a concentric hexagon shape for low-power radar interference and to an unequally-spaced pulse amplitude modulation (PAM) shape for the high-power counterpart.
\subsection{Dual-functional Radar-Communication (DFRC) System}
\emph{1) Information theory for the DFRC}
\\\indent It is well-understood that the radar works in a way that is fundamentally different from classic communication systems. Specifically, the communication takes place between two or more cooperative transceivers. By contrast, radar systems send probing signals to uncooperative targets, and infer useful information contained in the target echoes. To some degree, the process of radar target probing may be deemed as similar to the communication channel estimation, with the probing waveforms acting as the pilot symbols. For designing a DFRC system, one can unify radar and communication principles by invoking information theory, which may reveal fundamental performance bounds of the dual-functional systems \cite{7279172}.
\\\indent In a communication system, the transmitted symbols are drawn from a discrete constellation that is known to both TX and RX, which enables the use of \emph{bit rate} as a performance metric for the communication. By contrast, the useful information for radar is not in the probing waveform but rather in the echo wave reflected by the target, which is however not drawn from a finite-cardinality alphabet \cite{6875553}. Drawing parallels from information theory, one way to measure the radar information rate is to view each resolution unit of the radar as a ``constellation point", as each unit can accommodate a distinguishable point-like target. In \cite{7131098}, the ``channel capacity" of the radar is defined as the number of distinguishable targets, which is the maximum information that can be contained in the echo wave.
\\\indent In addition to the above definition, the authors of \cite{7279172} have considered the mutual information between the radar and the target. Intuitively, the variance of the noise imposed on the echo wave represents the uncertainty of the target information, and can be measured by the entropy of the echo. From an information theoretical viewpoint, the radar cancels part of the uncertainty by estimating the target parameters, where the remaining part is lower-bounded by the Cram\'er-Rao Bound (CRB), which can be viewed as the minimum variance achievable of the estimated parameter \cite{kay1998fundamentals}. In light of this methodology, the authors of \cite{7279172} consider a single-antenna DFRC receiver, which can process the target echo wave and the UL communication signal simultaneously. Such a channel can be viewed as a special multi-access (MAC) channel, where the target is considered as a virtual communication user. An estimation rate is defined as the information metric for the radar in \cite{7279172}. By invoking the analytical framework of the communication-only MAC channel, the trade-off between radar and communication performances is analyzed under different multi-access strategies. In \cite{chiriyath2017radar}, an integrated metric is proposed for the DFRC receiver, which is the weighted sum of the estimation and communication rates. More recently, this approach has been generalized to the multi-antenna DFRC system in \cite{8448783}. While the performance bounds of the DFRC systems have been specified by the above contributions, the design of DFRC waveforms is still an open problem.
\\\indent \emph{2) Temporal and spectral processing}
\\\indent In the early 1960s, the pioneering treatise \cite{mealey1963method} proposed to modulate communication bits onto radar pulses by the classical pulse interval modulation (PIM), which shows that one can design dual-functional waveforms by embedding useful information into radar signals. By realizing this, the authors of \cite{roberton2003integrated,saddik2007ultra} proposed to modulate chirp signals with communication bit sequences, where 0 and 1 are differentiated by exploiting the quasi-orthogonality of the up and down chirp waveforms. Likewise, the pseudo-random codes can also be used both as the probing signal and the information carrier \cite{jamil2008integrated}. A simpler approach is proposed in \cite{han2013joint} under a time-division framework, where the radar and the communication signals are transmitted in different time slots and thus do not interfere with each other.
\\\indent In addition to the above approaches where the DFRC waveforms are designed from the ground-up, a more convenient option would be to employ the existing communication signals for target detection. In this spirit, the classic Orthogonal Frequency Division Multiplexing (OFDM) signal is considered as a promising candidate \cite{garmatyuk2011multifunctional}. In \cite{sturm2011waveform}, the authors proposed to transmit OFDM communication signals for vehicle detection. The impact of the random data can be eliminated by simple element-wise division between the transmitted OFDM symbols and the received echoes. In contrast to its single-carrier counterpart, the OFDM approach of \cite{sturm2011waveform} employs the fast Fourier transform (FFT) and the inverse FFT (IFFT) for Doppler and range processing, respectively, which obtains the velocity and the range parameters in a decoupled manner. It is also possible to replace the sinusoidal subcarrier in the OFDM as the chirp signal \cite{7485314}. Accordingly, the fractional Fourier transform (FrFT) \cite{330368}, which is built upon orthogonal chirp basis, is used to process the target return.
\\\indent The above contributions have mainly investigated temporal and spectral processing for designing DFRC waveforms, while paying little information to the beneficial aspects of spatial processing. In what follows, we review the research progress in the design of MIMO DFRC systems.
\\\indent \emph{3) Spatial processing}
\\\indent Inspired by the space-division multiple access (SDMA) concept of MIMO communications, a straightforward MIMO DFRC scheme is to detect the target in the mainlobe of the radar antenna array, while transmitting useful information in the sidelobe. One can simply modulate the sidelobe level using amplitude shift keying (ASK), where different powers represent different communication symbols \cite{7347464}. Similarly, classic phase shift keying (PSK) could also be applied for representing the bits as the phases of the signals received at the angle of the sidelobe \cite{7485066}. Accordingly, multi-user communication can be implemented by varying the sidelobes at multiple angles. To avoid any undue performance-loss of the radar, beampattern invariance based approaches have been studied in \cite{7485316}, where the communication symbols are embedded by shuffling the transmitted waveforms across the antenna array. In this case, the information is embedded into the permutation matrices.
\\\indent In the above methods, a communication symbol is usually embedded into either a single or several radar pulses, which results in a low data rate that is tied to the pulse repetition frequency (PRF) of the radar, hence it is limited to the order of kbps. Moreover, the sidelobe embedding schemes can only work when the communication receiver benefits from a line-of-sight (LoS) channel. This is because for a multi-path channel, the received symbol will be seriously distorted by the dispersed signals arriving from Non-LoS (NLoS) paths, where all the sidelobe and the mainlobe power may contribute. To this end, the authors of \cite{8288677} proposed several beamforming designs to enable joint MIMO radar transmission and MU-MIMO communication, in which the communication signal was exploited for target detection, hence it would not affect the DL data rate. The joint beamforming matrix is optimized to approach an ideal radar beampattern, while guaranteeing the DL SINR and the power budget. To conceive the constant-modulus (CM) waveform design for DFRC systems, the recent contributions \cite{8386661,sspd_2019} proposed to minimize the DL multi-user interference (MUI) subject to specific radar waveform similarity and CM constraints. An efficient branch-and-bound (BnB) algorithm has been designed for solving the non-convex optimization problem, which finds the global optimum in tens of iterations.
\subsection{Limitations of the Existing Works}
Although there is a rich literature on various aspects of both RCC and DFRC scenarios, prior research mainly considers the applications in sub-6GHz band. To address the explosive growth of wireless devices and services, the forthcoming 5G network aims at an ambitious 1000-fold increase in capacity by exploiting the large bandwidth available in the mmWave band. In the meantime, it is expected that the mmWave BS will be equipped with beneficial sensing capability, which may find employment in a variety of scenarios such as vehicle-to-everything (V2X) communications. Dual-functional radar-communication in mmWave systems is a new and promising research area. Recent treatises \cite{8057284,8114253,8309274} propose to invoke the radar function to support V2X communications based on the IEEE 802.11ad WLAN protocol, which operates in the 60GHz band. As the WLAN standard is typically indoor based and employs small-scale antenna arrays, it can only support short-range sensing at the order of tens of meters. To overcome these drawbacks, the large-scale antenna arrays have to be exploited, which can compensate the high path-loss imposed on mmWave signals. Moreover, the high DoFs of massive antennas make it viable to support joint sensing and communication tasks. In order to reduce the hardware complexity and the associated costs, the maturing hybrid analog-digital (HAD) beamforming structure is typically used in such systems \cite{7010533,8030501}, which requires much fewer RF chains than the fully digital transceivers. While the authors of \cite{8550811} have presented analog beamforming designs for small-scale MIMO DFRC, little attention has been paid to HAD based massive MIMO (mMIMO) DFRC systems, which might be more practical, whilst maintaining compatibility with 5G mmWave applications.
\subsection{Main Contributions of Our Work}
In this paper, we propose a novel architecture for a DFRC system operating in the mmWave band. We consider a mMIMO mmWave BS that serves a multi-antenna UE while detecting multiple targets, where part of the targets are also the scatterers in the communication channel. To reduce the number of RF chains, an HAD beamformer is employed for both transmission and reception at the BS. We propose a novel DFRC frame structure that complies with state-of-the-art time-division duplex (TDD) protocols, which can be split into three stages for unifying similar radar and communication operations, namely 1) radar target search and communication channel estimation, 2) radar transmit beamforming and downlink communication and 3) radar target tracking and uplink communication. In each stage, we propose joint signal processing approaches that can fulfill both target detection and communication tasks via invoking hybrid beamforming. To be specific, in Stage 1, we estimate the AoAs of all the potential targets and the communication channel parameters by using both DL and UL pilots. Based on the estimation results, we propose in Stage 2 a novel joint HAD transmit beamforming design that can formulate directional beams towards the angles of interest, while equalizing the communication channel. Finally, in Stage 3 we track the angular variation by simultaneously processing the echoes of the targets while decoding the UL signal transmitted from the UE. Below we boldly and crisply summarize our contributions:
\begin{itemize}
  \item A novel mmWave mMIMO DFRC architecture that can simultaneously detect targets while communicating with the UE;
  \item A novel TDD frame structure capable of unifying radar and communication operations;
  \item A joint signal processing strategy that can search for unknown targets while estimating the communication channel;
  \item A joint HAD beamforming design that formulates directional beams towards targets of interest while equalizing the influence of the channel;
  \item A joint receiver design that can simultaneously track the variation of the targets while decoding the UL communication signals.
\end{itemize}
The remainder of this paper is organized as follows. Section III introduces the system model, Section IV proposes the basic framework of the DFRC system, Sections V-VII consider the signal processing schemes for Stages 1, 2 and 3, respectively, Section VIII provides numerical results. Finally, Section IX concludes the paper and identifies a number of future research directions.
\\\indent {\emph{Notation}}: Unless otherwise specified, matrices are denoted by bold uppercase letters (i.e., $\mathbf{H}$), vectors are represented by bold lowercase letters (i.e., $\bm{\alpha}$), and scalars are denoted by normal font (i.e., $\theta$). Subscripts indicate the location of the entry in the matrices or vectors (i.e., $\mathbf{F}_{RF}\left({i,j}\right)$ denotes the $(i,j)$th entry of $\mathbf{F}_{RF}$, $\mathbf{F}_{RF}\left({i,:}\right)$ and $\mathbf{F}_{RF}\left({:,j}\right)$ denote the \emph{i}th row and the \emph{j}th column of $\mathbf{F}_{RF}$, respectively). $\operatorname{tr}\left(\cdot\right)$, $\left(\cdot\right)^T$, $\left(\cdot\right)^H$, $\left(\cdot\right)^*$ and $\left(\cdot\right)^\dag$ stand for trace, transpose, Hermitian transpose, complex conjugate and pseudo-inverse, respectively. $\left\| \cdot\right\|$ and $\left\| \cdot\right\|_F$ denote the $l_2$ norm and the Frobenius norm respectively.

\section{System Model}
We consider an $N_t$-antenna massive MIMO DFRC BS that communicates with an $N_r$-antenna UE while detecting multiple targets. The system operates in TDD mode, and both BS and UE are assumed to be equipped with uniform linear arrays (ULA). To reduce the number of RF chains, the BS employs a fully-connected hybrid analog-digital beamforming structure with $N_{RF}$ RF chains, where $N_{RF} \le N_t$. Since the size of the antenna array at the UE is typically much smaller than at the BS, we assume that the UE adopts fully digital beamforming structure.
\begin{figure}
    \centering
    \includegraphics[width=0.9\columnwidth]{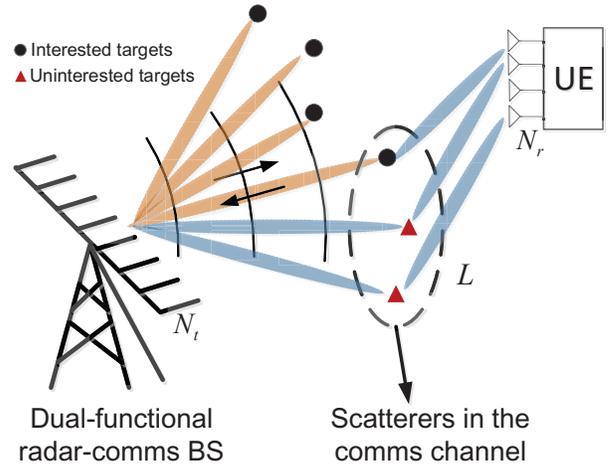}
    \caption{MmWave dual-functional radar-communication scenario.}
    \label{fig:1}
\end{figure}
\\\indent We show a generic DFRC scenario in Fig. 1, where a collection of $K$ scatterers/radar targets are randomly distributed within the communication/sensing environment, which are yet to be detected by the BS. While all targets reflect back the echo wave to the BS, not all of them contribute to communication scattering paths between the BS and the UE. Recent literature on mmWave channel modeling has shown that the scattering model describes well the mmWave communication channel, which typically has a small number of scattered paths. We assume that only $L$ out of $K$ scatterers are resolvable in the communication channel, and that $L \le N_r \le N_t$. Therefore, the rank of the communication channel is $L$, which suggests that the channel can support up to $L$ independent data streams to be transmitted simultaneously. For convenience, both $K$ and $L$ are assumed to be known to the BS.
\\\indent \emph{Remark 1:} From a radar perspective, not all targets are of interest. Obstacles such as trees and buildings, are unwanted reflectors and are commonly referred to as ``clutter" in the radar literature. Clutter interference can be avoided by not radiating or receiving in the corresponding directions. However, some of the clutter might come from significant scatterers in the communication channel (as shown by red triangles in Fig. 1). Therefore, for the purpose of estimating the channel parameters, it might still be necessary to beamform towards those scatterers. This is distinctly different from a pure radar target detection scenario. For convenience, we will not distinguish these two types of targets, and only identify the communication paths within the collection of all the targets, which will be discussed in detail in Sec. V.
\\\indent \emph{Remark 2:} There might also exist targets that are neither significant scatterers in the communication channel nor of any interest to the DFRC BS. For notational convenience and following most of the seminal literature in the area \cite{7814210,7089157,7470514,7898445,8355705,4655353}, we will not discuss such targets in detail and simply incorporate the generated interference in the noise term.
\\\indent While the hybrid beamforming technique is popular in mmWave mMIMO communications, it can be useful in the radar area as well. In fact, the HAD structure has already been exploited to design a type of novel radar system referred to as ``phased-MIMO radar" \cite{5419124}, which is a compromise between the phased-array radar and the MIMO radar. There has been a long debate in the radar community on which type of radar has better performance since the birth of the MIMO radar concept in 2004 \cite{1316398,5109947}. To be specific, the MIMO radar transmits independent waveforms by each antenna by employing a fully digital beamformer, whereas the phased-array radar transmits via each antenna the phase-shifted counterpart of a benchmark signal, which indicates that there are multiple phase shifters and antennas but only a single RF chain used by the phased-array radar. By exploiting higher degrees-of-freedom (DoFs) and waveform diversity, the MIMO radar achieves higher detection probability at the cost of increasing the computational overhead and the hardware complexity \cite{4350230,li2008mimo}. Moreover, due to the non-coherent combination of the received signals, the receive SINR of the MIMO radar is lower than that of its phased-array counterpart \cite{5419124}. It is against this background that the phased-MIMO radar has been proposed. By partitioning the antenna array into several sub-arrays \cite{6104178}, the phased-MIMO radar transmits individual digital signals by each RF chain, but performs coherent analog combination at each sub-array, which is expected to strike a favorable performance tradeoff between both types of radars. Given the similarities between the HAD communication and the phased-MIMO radar, we consider their combination in the proposed DFRC system.

\subsection{Radar Model}
Let $\mathbf{X}_r\in {\mathbb{C}}^{N_t \times T}$ be a probing signal matrix sent by the BS, which is composed by $T$ snapshots along the fast-time axis. The echo wave reflected by the targets received at the BS can be expressed as
\begin{equation}\label{eq1}
{{\mathbf{Y}}_{echo}} = \sum\limits_{k = 1}^K {{\alpha _k}{\mathbf{a}}\left( {{\theta _k}} \right)} {{\mathbf{a}}^T}\left( {{\theta _k}} \right){{\mathbf{X}}_r} + {\mathbf{Z}},
\end{equation}
where ${\alpha _k}$ denotes the complex-valued reflection coefficient of the \emph{k}th target, $\theta_k$ is the \emph{k}th target's azimuth angle, with ${{\mathbf{a}}\left( {{\theta}} \right)}$ being the steering vector of the transmit antenna array, finally ${\mathbf{Z}} \in {\mathbb{C}}^{N_t \times T}$ represents the noise plus interference, with the variance $\sigma_r^2$. In the case of ULA, the steering vector can be written in the form
\begin{equation}\label{eq2}
  {\mathbf{a}}\left( \theta  \right) = {\left[ {1,{e^{j\frac{{2\pi }}{\lambda }d\sin \left( \theta  \right)}},...,{e^{j\frac{{2\pi }}{\lambda }d\left( {{N_t} - 1} \right)\sin \left( \theta  \right)}}} \right]^T} \in {\mathbb{C}}^{N_t \times 1},
\end{equation}
where $d$ and $\lambda$ denote the antenna spacing and the signal wavelength. Without loss of generality, we set $d = \lambda/2$. Following the standard assumptions in the literature \cite{4276989,4655353,li2008mimo}, the signal model in (\ref{eq1}) is assumed to be obtained in a particular range-Doppler bin of interest, for which the range and the Doppler parameters are omitted in the model.
\\\indent By arranging the steering vectors into a steering matrix ${\mathbf{{A}}}\left( \Theta  \right) = \left[ {{\mathbf{a}}\left( {{\theta _1}} \right),...,{\mathbf{a}}\left( {{\theta _K}} \right)} \right]$, the reflected signal model in (\ref{eq1}) can be equivalently recast as
\begin{equation}\label{eq3}
{{\mathbf{Y}}_{echo}} = {\mathbf{{A}}}\left( \Theta  \right)\operatorname{diag} \left( {\bm{\alpha }} \right){{\mathbf{{A}}}^T}\left( \Theta  \right){{\mathbf{X}}_r} + {\mathbf{Z}},
\end{equation}
where ${\bm{\alpha }} = \left[ \alpha_1,...,\alpha_K\right ]^T$, $\Theta  = \left\{ {{\theta _1},{\theta _2},...,{\theta _K}} \right\}$.

\subsection{MmWave Communication Model}
Let $\mathbf{X}_{DL} \in {\mathbb{C}}^{N_t \times T}$ be a DL signal matrix sent from the BS to the UE, the received signal model at the UE can be formulated as
\begin{equation}\label{eq4}
  {{\mathbf{Y}}_{DL}} = {\mathbf{H}}{{\mathbf{X}}_{DL}} + {\mathbf{N}_{DL}},
\end{equation}
where ${\mathbf{N}_{DL}}\in {\mathbb{C}}^{N_r \times T}$ denotes the noise with the variance of $\sigma_{DL}^2$, and ${\mathbf{H}}\in {\mathbb{C}}^{N_r \times N_t}$ is the narrowband mmWave communication channel, which is assumed constant throughout the duration $T$, and can be expressed as follows by use of the extended Saleh-Valenzuela model \cite{8052157,7888145}
\begin{equation}\label{eq5}
  {\mathbf{H}} = \sum\limits_{l = 1}^L {{\beta _l}{\mathbf{b}}} \left( {{\phi _l}} \right){{\mathbf{a}}^T}\left( {{\varphi _l}} \right),
\end{equation}
where $\beta _l, \phi _l$ and $\varphi _l$ denote the complex scattering coefficient, the Angle of Arrival (AoA) and the Angle of Departure (AoD) of the \emph{l}th scattering path, and
\begin{equation}\label{eq6}
  {\mathbf{b}}\left( \phi  \right) = {\left[ {1,{e^{j\frac{{2\pi }}{\lambda }d\sin \left( \phi  \right)}},...,{e^{j\frac{{2\pi }}{\lambda }d\left( {{N_r} - 1} \right)\sin \left( \phi  \right)}}} \right]^T}\in {\mathbb{C}}^{N_r \times 1}
\end{equation}
represents the steering vector of the UE's antenna array. Note that the scatterers of the communication channel are also part of the targets being detected by the BS. From the perspective of the UE, the AoDs $\varphi_l,\forall l$ belong to the set of AoAs $\Theta  = \left\{ {{\theta _1},...,{\theta _K}} \right\}$ of radar targets seen from the BS. We assume, without loss of generality, that $\varphi_l = \theta_l, l = 1,...,L$. The received signal can be therefore re-arranged as
\begin{equation}\label{eq32}
  {{\mathbf{Y}}_{DL}} = {\mathbf{B}}\left( \Phi  \right)\operatorname{diag} \left( {\bm{\beta }} \right){{\mathbf{A}}^T}\left( {{\Theta _1}} \right){{\mathbf{X}}_{DL}} + {\mathbf{N}}_{DL},
\end{equation}
where
\begin{equation}\label{eq33}
\begin{gathered}
  {\mathbf{B}}\left( \Phi  \right) = \left[ {{\mathbf{b}}\left( {{\phi _1}} \right),...,{\mathbf{b}}\left( {{\phi _L}} \right)} \right],{\mathbf{A}}\left( {{\Theta _1}} \right) = \left[ {{\mathbf{a}}\left( {{\theta _1}} \right),...,{\mathbf{a}}\left( {{\theta _L}} \right)} \right] \hfill \\
  {\mathbf{\beta }} = {\left[ {{\beta _1},...,{\beta _L}} \right]^T},\Phi  = \left[ {{\phi _1},...,{\phi _L}} \right],{\Theta _1} = \left\{ {{\theta _1},...,{\theta _L}} \right\} \subseteq \Theta.  \hfill \\
\end{gathered}
\end{equation}
Given the reciprocity of the TDD channel, the UL communication model can be accordingly expressed as
\begin{equation}\label{eq9-1}
  {{\mathbf{Y}}_{UL}} = {\mathbf{A}}\left( {{\Theta _1}} \right)\operatorname{diag} \left( {\bm \beta}  \right){{\mathbf{B}}^T}\left( \Phi  \right){{\mathbf{X}}_{UL}} + {{\mathbf{N}}_{UL}},
\end{equation}
where $\mathbf{X}_{UL} \in {\mathbb{C}}^{N_r \times T}$ denotes the UL communication signal, and ${\mathbf{N}_{UL}}\in {\mathbb{C}}^{N_t \times T}$ represents the noise having the variance of $\sigma_{UL}^2$.
\\\indent It can be observed in the model of (\ref{eq5}) and (\ref{eq9-1}) that the mmWave communication channel has an intrinsic geometric structure, which makes it equivalent to a bi-static radar channel \cite{6994289}, where the radar's TX and RX antennas are widely separated instead of being collocated as in the mono-static case of (\ref{eq1}). Accordingly, the scatterers act as known or unknown radar targets, depending on whether the channel has been estimated. Note that such equivalences do not hold for channels modeled by stochastic distributions, e.g., Rayleigh distribution, which contain little information about the geometric environment over which the communication takes place.

\begin{figure*}[!t]
    \centering
    \includegraphics[width=1.8\columnwidth]{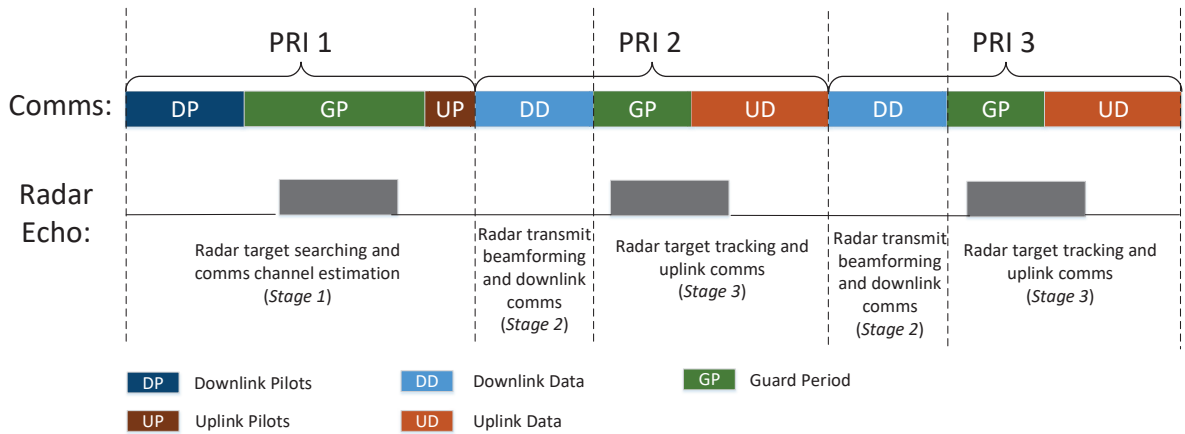}
    \caption{Frame structure of the DFRC system.}
    \label{fig:2}
\end{figure*}
\section{The Dual-functional Radar-Communication Framework}
We further reveal some important insights by taking a closer look at both the radar and the communication models.
\\\indent \emph{Remark 3:} The aim of the communication is to decode data from the noisy signal under the knowledge of the channel state information. On the other hand, the radar acquires the geometric information of targets by sending a known probing signal. This indicates that, radar target detection is more similar to the channel estimation process rather than to the data communication itself.
\\\indent \emph{Remark 4:} Radar detection can also be viewed as a special communication scenario, where the targets unwillingly transmit their geometric information to the radar. Therefore, the radar targets may act as virtual communication users that communicate with the radar in an uncooperative manner.
\\\indent Inspired by the above remarks, we propose the following mmWave DFRC framework, which aims for unifying radar and communication operations by joint signal processing, and can be generally split into the following three stages:
\\\indent \emph{1) Radar target search and communication channel estimation}
\\\indent When the radar has no {\emph{a priori}} knowledge about targets, the initial step is to search for potential targets in the whole angular domain. Similarly, when no channel information is available at the communication system, the CSI has to be estimated before any useful information can be decoded at the receiver. Note that both operations require a signal with beneficial auto- and cross-correlation properties in order to extract the target parameters or the scattering characteristics of the channel. Hence, it is natural to combine the two operations into a joint process. More specifically, in our case, the BS first sends omnidirectional DL pilots (DP), and then estimates the AoAs of all \emph{K} targets in $\Theta$. The UE also receives the probing waveform through \emph{L} scattering paths, based on which it estimates $L$ AoDs in $\Phi$, and sends back UL pilots (UP) to the BS. By exploiting the reciprocity of the DL and the UL channels, the BS is able to identify those targets which also play the role of scatterers in the communication link. We propose a joint solution for this operation in Sec. V.
\\\indent \emph{2) Radar transmit beamforming and downlink communication}
\\\indent After the first stage, the BS will have the estimate of $\Theta$ for all the targets. Nevertheless, the estimate of $\Phi$ is only available at the UE. The BS then formulates directional DL beams towards the angles of the targets of interest by designing a joint sensing-communication beamformer, and obtains more accurate observations. In the meantime, the joint beamformer designed aims for pre-equalizing the communication channel effects, so that the data can be correctly decoded at the UE. We propose and detail a joint solution for this operation in Sec. VI.
\\\indent \emph{3) Radar target tracking and uplink communication}
\\\indent After Stage 2, the BS may receive both the target echoes and the UL signals, based on which it tracks the variation of target parameters while decoding the UL data transmitted from the UE. As we have discussed above, the targets can be viewed as virtual UEs that passively transmit their geometric parameters to the BS by reflecting the probing signal. In this spirit, we design sophisticated receive signal processing approaches to jointly fulfill both requirements, i.e., target parameter estimation and data decoding. We propose and detail a joint solution for this operation in Sec. VII.
\\\indent As shown in Fig. 2, a specifically tailored frame structure is designed to coordinate the above DFRC operations based on a typical TDD protocol. In Stage 1, the BS transmits omnidirectional waveforms to search for targets and to estimate the communication channel, and then receives both the echoes from the targets and the UP from the UE. Since all the targets/scatterers are distributed in between the BS and the UE, and that the echoes are reflected instantaneously after hitting the targets, the round-trip from the BS to the targets/scatterers is typically shorter than that from the BS to the UE given the processing delay of the UL communication. For this reason, we assume that the target echoes are always received ahead of the UL transmission. It is worth noting that a guard period\footnote{Note that the GP is typically used in TDD protocols such as TDD-LTE.} (GP) is required between DP and UP to avoid the interference between UP and target echoes \cite{7504377}. The GP should be long enough to cover the longest round-trip delay plus the length of the DP. In addition, the UP is designed to be much shorter than the DP to further avoid collision of the received signals. In Stage 2, the BS transmits DL data while formulating directional beams towards all the directions in $\Theta$ based on the measurements in Stage 1. In Stage 3, the BS receives both the echoes and the UL data, based on which it tracks the variation of the targets while decoding the UL information. Here we reserve a shorter GP between DL and UL operations to guarantee a high UL data rate. As the UL data sequences are much longer than the UP in Stage 1, collision between the echo wave and the data is inevitable. To this end, we propose a successive interference cancellation (SIC) approach \cite{tse2005fundamentals} at Stage 3 to mitigate the interference from the targets, which will be discussed in Sec. VII. It can be noted from above that the BS indeed acts as a pulsing radar that repeatedly transmits pulses and receives both echoes and UL signals. Following the standard radar literature, we term a transmit-receive cycle as a \emph{pulse repetition interval (PRI)}.
\\\indent In what follows, we will design signal processing strategies for the above three stages, respectively.

\section{Stage 1: Radar Target Search and Communication Channel Estimation}
In this section, we first introduce a novel pilot signal generation method for the purpose of joint target search and CSI acquisition, and then propose parameter estimation approaches at both the BS and the UE.
\subsection{Pilot Signal Generation Using Hybrid Structure}
Given a DP signal matrix ${\mathbf{S}}_{DP} \in {\mathbb{C}^{N_t \times T}}$, it is well-known in the field of channel estimation that the optimal performance can be achieved if its covariance matrix satisfies
\begin{equation}\label{eq7}
  {{\mathbf{R}}_s} = \frac{1}{T}{{\mathbf{S}}_{DP}}{{{\mathbf{S}}_{DP}^H}} = \frac{P}{{{N_t}}}{{\mathbf{I}}_{{N_t}}},
\end{equation}
where $P$ is the total transmit power. It can be seen from above that the optimal pilot signal transmitted on each antenna should be spatially orthogonal. Similar investigations in the MIMO radar literature have also revealed that, the CRB of target parameter estimation can be minimized by the use of orthogonal waveforms \cite{li2008mimo}, in which case the spatial beampattern can be written as
\begin{equation}\label{eq8}
  d\left( \theta  \right) = {{\mathbf{a}}^T}\left( \theta  \right){{\mathbf{R}}_s}{\mathbf{a}}^*\left( \theta  \right) = P, \;\;\forall \theta,
\end{equation}
which is an omnidirectional beampattern. Naturally, such a beampattern transmits equivalent power at each angle, and will hence search for targets over the whole angular domain.
\\\indent At a first glance, it seems that any orthogonal waveform can be used for both radar target search and channel estimation. Nevertheless, there are still some radar-specific requirements that the probing waveform should satisfy. For instance, waveforms having large time-bandwidth product (TBP) are preferred by the radar, as it offers performance improvement in both the range resolution and the maximum detectable range. To this end, we propose to employ orthogonal linear frequency modulation (LFM) signals, which are commonly used MIMO radar waveforms. According to \cite{7450660}, the $\left(n,t\right)$th entry of a orthogonal LFM waveform matrix can be defined as
\begin{equation}\label{eq9}
  {{\mathbf{S}}_{DP}}\left( {n,t} \right) = \sqrt {\frac{P}{{{N_t}}}} \exp \left( {\frac{{j2\pi n\left( {t - 1} \right)}}{T}} \right)\exp \left( {\frac{{j\pi {{\left( {t - 1} \right)}^2}}}{T}} \right).
\end{equation}
It can be readily proven that (\ref{eq9}) satisfies the orthogonality property (\ref{eq7}). Next, we consider to generate such a waveform matrix by invoking the HAD array. Let us denote the baseband signal matrix by ${\mathbf{S}}_{BB} \in {\mathbb{C}^{N_{RF} \times T}}$, and the analog precoding matrix with unit-modulus entries by ${\mathbf{F}}_{RF} \in {\mathbb{C}^{N_{t} \times N_{RF}}}$. The problem is to design both ${\mathbf{F}}_{RF}$ and ${\mathbf{S}}_{BB}$, such that
\begin{equation}\label{eq10}
  {\mathbf{F}}_{RF}{\mathbf{S}}_{BB} = {{\mathbf{S}}_{DP}}.
\end{equation}
Due to the non-convex unit-modulus constraints imposed on ${\mathbf{F}}_{RF}$, it is difficult to solve the above equation directly. We therefore propose a construction method in the following.
\\\indent For the signal transmitted on the \emph{n}th antenna, note that as per (\ref{eq9}), the following equation holds true for any adjacent time-slots
\begin{equation}\label{eq11}
   \frac{{{{\mathbf{S}}_{DP}}\left( {n,t + 1} \right)}}{{{{\mathbf{S}}_{DP}}\left( {n,t} \right)}} = \exp \left( {\frac{{j2\pi n}}{T}} \right)\exp \left( {\frac{{j\pi \left( {2t - 1} \right)}}{T}} \right).
\end{equation}
By introducing the notation of
\begin{equation}\label{eq12}
  {u_n} = \exp \left( {\frac{{j2\pi n}}{T}} \right),{v_t} = \exp \left( {\frac{{j\pi \left( {2t - 1} \right)}}{T}} \right),
\end{equation}
\begin{equation}\label{eq13}
  {\mathbf{u}} = {\left[ {{u_1},{u_2},...,{u_{{N_t}}}} \right]^T},
\end{equation}
it follows that
\begin{equation}\label{eq14}
  {{\mathbf{S}}_{DP}}\left( {:,t + 1} \right) = \operatorname{diag} \left( {\mathbf{u}} \right){{\mathbf{S}}_{DP}}\left( {:,t} \right){v_t},\forall t,
\end{equation}
where ${{\mathbf{S}}_{DP}}\left( {:,t} \right)$ denotes the \emph{t}th column of ${{\mathbf{S}}_{DP}}$.
\\\indent In order to generate ${{\mathbf{S}}_{DP}}$, we consider a simple strategy where the analog beamforming matrix changes on a time-slot basis, in which case the following two equations should be satisfied
\begin{subequations}\label{eq15}
\begin{align}
  & {\mathbf{F}}_{RF}^1{{\mathbf{S}}_{BB}}\left( {:,1} \right) = {{\mathbf{S}}_0}\left( {:,1} \right), \hfill \\
  & {\mathbf{F}}_{RF}^{t + 1}{{\mathbf{S}}_{BB}}\left( {:,t + 1} \right) = \operatorname{diag} \left( {\mathbf{u}} \right){\mathbf{F}}_{RF}^t{{\mathbf{S}}_{BB}}\left( {:,t} \right){v_t},
\end{align}
\end{subequations}
where ${\mathbf{F}}_{RF}^t$ denotes the analog beamforming matrix at the \emph{t}th time-slot. Therefore, it is sufficient to let
\begin{subequations}\label{eq16}
\begin{align}
  & {\mathbf{F}}_{RF}^{t + 1} = \operatorname{diag} \left( {\mathbf{u}} \right){\mathbf{F}}_{RF}^t, \forall t,\hfill \\
  & {{\mathbf{S}}_{BB}}\left( {:,t + 1} \right) = {{\mathbf{S}}_{BB}}\left( {:,t} \right){v_t},\forall t.
\end{align}
\end{subequations}
Furthermore, noting that ${{\mathbf{S}}_{DP}}\left( {:,1} \right) = \sqrt {P/{N_t}} \mathbf{1}_{N_t}$, we can simply choose
\begin{equation}\label{eq17}
{\mathbf{F}}_{RF}^1 = {{\mathbf{1}}_{{N_t}}}{\mathbf{1}}_{{N_{RF}}}^T,{{\mathbf{S}}_{BB}}\left( {:,1} \right) = \sqrt {\frac{P}{{N_{RF}^2{N_t}}}} {{\mathbf{1}}_{{N_{RF}}}}, \forall t,
\end{equation}
where ${{\mathbf{1}}_{{N}}}$ denotes the $N \times 1$ all-one vector. By the above method, the analog beamforming matrix and the baseband signal can be generated at each time-slot in a recursive manner. One can thus generate the LFM waveform in (\ref{eq9}) for target search and channel estimation.

\subsection{Parameter Estimation}
After transmitting the waveform ${\mathbf{S}}_{DP}$ using the HAD architecture, the BS receives the signals reflected from the targets, which can be expressed as
\begin{equation}\label{eq20}
{{\mathbf{Y}}_{echo}} = \sum\limits_{k = 1}^K {{\alpha _k}{\mathbf{a}}\left( {{\theta _k}} \right)} {{\mathbf{a}}^T}\left( {{\theta _k}} \right){{\mathbf{S}}_{DP}} + {\mathbf{Z}}.
\end{equation}
Then, the signal after analog combination can be accordingly expressed by
\begin{equation}\label{eq21}
\begin{gathered}
  {{{\mathbf{\tilde Y}}}_{echo}} = \sum\limits_{k = 1}^K {{\alpha _k}{{\mathbf{W}}_{RF}}{\mathbf{a}}\left( {{\theta _k}} \right)} {{\mathbf{a}}^T}\left( {{\theta _k}} \right){{\mathbf{S}}_{DP}} + {{\mathbf{W}}_{RF}}{\mathbf{Z}} \hfill \\
   = \sum\limits_{k = 1}^K {{\alpha _k}{\mathbf{\tilde a}}\left( {{\theta _k}} \right)} {{\mathbf{a}}^T}\left( {{\theta _k}} \right){{\mathbf{S}}_{DP}} + {\mathbf{\tilde Z}}, \hfill \\
\end{gathered}
\end{equation}
where ${{\mathbf{W}}_{RF}} \in {\mathbb{C}^{{N_{RF}} \times {N_t}}}$ is the analog combination matrix having unit-modulus entries, ${\mathbf{\tilde a}}\left( \theta  \right) = {{\mathbf{W}}_{RF}}{\mathbf{a}}\left( \theta  \right) \in {\mathbb{C}^{{N_{RF}} \times 1}}$ is the equivalent receive steering vector, and ${\mathbf{\tilde Z}} = {{\mathbf{W}}_{RF}}{\mathbf{Z}}$. Since no {\emph{a priori}} knowledge about the AoAs is available at this stage, there is no preference on the choice of the analog beamformer. To this end, we assume that each entry of ${{\mathbf{W}}_{RF}}$ is randomly drawn from the unit circle.
\\\indent To estimate the angles, we invoke the classic MUltiple SIgnal Classification (MUSIC) algorithm, which is known to have high angle resolution \cite{1143830}. Nevertheless, the conventional MUSIC approach requires the processing of multiple receptions of the reflected pulses. Typically, the number of such observations should be larger than the size of the antenna array, which is not realistic in the case of massive MIMO. Therefore, we propose a modification of the MUSIC algorithm where we estimate the AoAs using a single pulse. Let ${\mathbf{\tilde A}}\left( \Theta  \right) = \left[ {{\mathbf{\tilde a}}\left( {{\theta _1}} \right),...,{\mathbf{\tilde a}}\left( {{\theta _K}} \right)} \right]$. The eq. (\ref{eq21}) can be equivalently recast as
\begin{equation}\label{eq22}
  {{{\mathbf{\tilde Y}}}_{echo}} = {\mathbf{\tilde A}}\left( \Theta  \right)\operatorname{diag} \left( {\bm{\alpha }} \right){{\mathbf{A}}^T}\left( \Theta  \right){{\mathbf{S}}_{DP}} + {\mathbf{\tilde Z}}.
\end{equation}
Note the fact that $\frac{1}{{{N_t}}}{{\mathbf{W}}_{RF}}{\mathbf{W}}_{RF}^H \approx {{\mathbf{I}}_{{N_{RF}}}}$ when $N_t$ is sufficiently large. By recalling (\ref{eq7}), the covariance matrix of (\ref{eq22}) is given by
\begin{equation}\label{eq23}
  \begin{gathered}
  {{\mathbf{R}}_{\tilde Y}} = \frac{1}{T}{{{\mathbf{\tilde Y}}}_{echo}}{\mathbf{\tilde Y}}_{echo}^H = \frac{P}{{{N_t}}}{\mathbf{\tilde A}}\left( \Theta  \right){{\mathbf{R}}_s}{{{\mathbf{\tilde A}}}^H}\left( \Theta  \right) + \frac{1}{T}{\mathbf{\tilde Z}}{{{\mathbf{\tilde Z}}}^H} \hfill \\
   = \frac{P}{{{N_t}}}{\mathbf{\tilde A}}\left( \Theta  \right){{\mathbf{R}}_s}{{{\mathbf{\tilde A}}}^H}\left( \Theta  \right) + \sigma _r^2{{\mathbf{W}}_{RF}}{\mathbf{W}}_{RF}^H \hfill \\
   \approx \frac{P}{{{N_t}}}{\mathbf{\tilde A}}\left( \Theta  \right){{\mathbf{R}}_s}{{{\mathbf{\tilde A}}}^H}\left( \Theta  \right) + \sigma _r^2{N_t}{{\mathbf{I}}_{{N_{RF}}}}, \hfill \\
\end{gathered}
\end{equation}
where ${{\mathbf{R}}_s} = \operatorname{diag} \left( {\bm{\alpha }} \right){{\mathbf{A}}^T}\left( \Theta  \right){{\mathbf{A}}^*}\left( \Theta  \right)\operatorname{diag} \left( {{{\bm{\alpha }}^*}} \right)$. Following the standard MUSIC algorithm, the eigenvalue decomposition of (\ref{eq23}) is formulated as
\begin{equation}\label{eq24}
  {{\mathbf{R}}_{\tilde Y}} = \left[ {{{\mathbf{U}}_s},{{\mathbf{U}}_n}} \right]\left[ {\begin{array}{*{20}{c}}
  {{{\mathbf{\Sigma }}_s}}&{} \\
  {}&{{{\mathbf{\Sigma }}_n}}
\end{array}} \right]\left[ \begin{gathered}
  {\mathbf{U}}_s^H \hfill \\
  {\mathbf{U}}_n^H \hfill \\
\end{gathered}  \right],
\end{equation}
where ${{\mathbf{U}}_s} \in \mathbb{C}^{{N}_{RF} \times K}$ and ${{\mathbf{U}}_n} \in \mathbb{C}^{{N}_{RF} \times {\left({N}_{RF}-K\right)}}$ contain eigenvectors, which span the signal and the noise subspaces, respectively. It then follows that
\begin{equation}\label{eq25}
  \operatorname{span} \left( {{\mathbf{\tilde A}}\left( \Theta  \right)} \right) = \operatorname{span} \left( {{{\mathbf{U}}_s}} \right),\operatorname{span} \left( {{\mathbf{\tilde A}}\left( \Theta  \right)} \right) \bot \operatorname{span} \left( {{{\mathbf{U}}_n}} \right),
\end{equation}
which suggests that ${\mathbf{\tilde a}}\left( \theta_k  \right), \forall k$ are orthogonal to ${{{\mathbf{U}}_n}}$. The MUSIC spectrum can be thus formulated as
\begin{equation}\label{eq26}
  {P_\text{MUSIC}}\left( \theta  \right) = \frac{1}{{{{{\mathbf{\tilde a}}}^H}\left( \theta  \right){{\mathbf{U}}_n}{\mathbf{U}}_n^H{\mathbf{\tilde a}}\left( \theta  \right)}}.
\end{equation}
By finding the \emph{K} largest peaks of (\ref{eq26}), we can readily locate the AoAs of the \emph{K} targets.
\\\indent The next step is to estimate $\alpha_k$ associated with each AoA. Since the estimated angle $\hat \theta_k$ is now available, we employ the Angle and Phase EStimation (APES) algorithm of \cite{4655353,506612} to obtain an estimated $\alpha_k$ with superior accuracy. Given each estimated $\hat \theta_k$, the APES technique aims at solving the following optimization problem \cite{4655353}
\begin{equation}\label{eq27}
\begin{gathered}
  \mathop {\min }\limits_{{\mathbf{w}}_k,{\alpha _k}} \;{\left\| {{{\mathbf{w}}_k^H}{{{\mathbf{\tilde Y}}}_{echo}} - {\alpha _k}{{\mathbf{a}}^T}\left( {{\hat \theta _k}} \right){{\mathbf{S}}_0}} \right\|^2} \hfill \\
  s.t.\;\;\;{{\mathbf{w}}_k^H}{\mathbf{\tilde a}}\left( \hat \theta  \right) = 1, \hfill \\
\end{gathered}
\end{equation}
where ${\mathbf{w}}_k \in \mathbb{C}^{{N}_{RF} \times 1}$ is the weighting vector. Following \cite{4655353}, the optimal ${\mathbf{w}}_k$ of (\ref{eq27}) can be expressed as
\begin{equation}\label{eq28}
  {{\mathbf{w}}_k} = \frac{{{{\mathbf{Q}}^{ - 1}}{\mathbf{\tilde a}}\left( \hat \theta_k  \right)}}{{{{{\mathbf{\tilde a}}}^H}\left( \hat \theta_k  \right){{\mathbf{Q}}^{ - 1}}{\mathbf{\tilde a}}\left( \hat \theta_k  \right)}},
\end{equation}
where
\begin{equation}\label{eq29}
{\mathbf{Q}} = {{\mathbf{R}}_{\tilde Y}} - \frac{1}{{{T^2}P}}{{{\mathbf{\tilde Y}}}_{echo}}{\mathbf{S}}_{DP}^H{{\mathbf{a}}^*}\left( {{\hat \theta _k}} \right){{\mathbf{a}}^T}\left( {{\hat \theta _k}} \right){{\mathbf{S}}_{DP}}{\mathbf{\tilde Y}}_{echo}^H.
\end{equation}
Accordingly, the \emph{k}th complex amplitude can be estimated by
\begin{equation}\label{eq30}
  {{\hat \alpha }_k} = \frac{1}{{TP}}{\mathbf{w}}_k^H{{{\mathbf{\tilde Y}}}_{echo}}{\mathbf{S}}_{DP}^H{{\mathbf{a}}^*}\left( {{\hat \theta _k}} \right).
\end{equation}
\\\indent We then estimate the angle parameters at the UE, where the received signal matrix at the UE can be formulated as
\begin{equation}\label{eq31}
  {{\mathbf{Y}}_{DL}} = \sum\limits_{l = 1}^L {{\beta _l}{\mathbf{b}}} \left( {{\phi _l}} \right){{\mathbf{a}}^T}\left( {{\varphi  _l}} \right){{\mathbf{S}}_{DP}} + {\mathbf{N}}_{DL}.
\end{equation}
Similarly, the UE estimates $\Phi$ by the MUSIC algorithm, and obtains the estimated angles $\hat \Phi  = \left\{ {{{\hat \phi }_1}},...,{{{\hat \phi }_L}} \right\}$.

\subsection{Identifying Communication Channel Paths from Targets}
While the BS has the knowledge of all the AoAs of targets, it still remains for us to distinguish which targets contribute to the scattering paths in the communication channel. In other words, the BS has to separate $\Theta_1$ from $\Theta_2 \triangleq \Theta \setminus \Theta_1$. Also, the BS still has to estimate $\beta_l$ for each scattering path, as this is not equivalent to the reflection coefficient $\alpha_l$.
\\\indent With the estimated $\hat \Phi$ at hand, the UE formulates the following transmit beamformer
\begin{equation}\label{eq34}
  {{\mathbf{F}}_{UE}} = {\mathbf{B}^*}\left( {\hat \Phi } \right){\left( {{{\mathbf{B}}^T}\left( {\hat \Phi } \right){\mathbf{B}^*}\left( {\hat \Phi } \right)} \right)^{ - 1}} \in {\mathbb{C}^{{N_r} \times L}},
\end{equation}
which aims for zero-forcing the steering matrix ${\mathbf{B}}\left( \Phi  \right)$. Following the frame structure proposed in Sec. IV, the UE then sends a very short UP to the BS by using ${{\mathbf{F}}_{UE}}$. Without loss of generality, we assume that the UP is an identity matrix $\mathbf{I}_L$. If $\Phi$ is perfectly estimated, the signal received at the BS is
\begin{equation}\label{eq35}
\begin{gathered}
  {{\mathbf{Y}}_{UL}} = {\mathbf{A}}\left( {{\Theta _1}} \right)\operatorname{diag} \left( {\bm{\beta }} \right){{\mathbf{B}}^T}\left( \Phi  \right){{\mathbf{F}}_{UE}}{{\mathbf{I}}_L} + {\mathbf{N}}_{UL} \hfill \\
   = {\mathbf{A}}\left( {{\Theta _1}} \right)\operatorname{diag} \left( {\bm{\beta }} \right) + {\mathbf{N}}_{UL}.\hfill \\
\end{gathered}
\end{equation}
To identify $\Theta_1$, the BS formulates the new analog combiner $\mathbf{G}_{{RF}} \in \mathbb{C}^{K \times N_{t}}$ with \emph{K} RF chains being activated, where the \emph{k}th row of $\mathbf{G}_{{RF}}$ is given as
\begin{equation}\label{eq36}
  {{\mathbf{G}}_{RF}}\left( {k,:} \right) = {{\mathbf{a}}^H}\left( {{\hat \theta _k}} \right),\forall k.
\end{equation}
After analog combination, the BS picks the specific $L$ entries having the $L$ largest moduli from ${{\mathbf{G}}_{RF}}{{\mathbf{y}}_{UL}}$, which are generated by the signal arriving from \emph{L} AoAs of the communication channel. By doing so, the BS can identify $\Theta_1$.
\\\indent Finally, we rebuild the steering matrix ${\mathbf{A}}\left( {{{\hat{\Theta}}_1}} \right)$, and estimate the scattering coefficients $\beta_l,\forall l$ by the simple least-squares (LS) estimation. For clarity, we summarize the target search/channel estimation process in Algorithm 1.
\begin{algorithm}
\caption{Stage 1: Radar Target Search and Communication Channel Estimation}
\label{alg:A}
\begin{algorithmic}
    \STATE \textbf{Step 1}: BS sends DP to search targets and estimate the channel.
    \STATE \textbf{Step 2}: BS receives the echo wave from targets, and estimates $\Theta$ and $\bm \alpha$ using MUSIC and APES.
    \STATE \textbf{Step 3}: UE receives DP, and estimates $\Phi$ using MUSIC.
    \STATE \textbf{Step 4}: UE formulates the ZF beamformer based on (\ref{eq34}), and transmits UP.
    \STATE \textbf{Step 5}: BS receives the UP, identifies the communication paths by the analog combiner (\ref{eq36}), and estimates $\bm \beta$ using the LS estimator.
\end{algorithmic}
\end{algorithm}
\section{Stage 2: Radar Transmit Beamforming and Downlink Communication}
In this section, we propose a novel joint transmit beamforming design at the BS for both target detection and DL communication by invoking the HAD structure. For supporting the radar functionality, we formulate directional beams towards the targets of interest to obtain more accurate observations. For the communication aspect, on the other hand, we equalize the channel.
\subsection{Problem Formulation}
Our goal is to design the analog and the digital beamforming matrices $\mathbf{F}_{RF} $ and $\mathbf{F}_{BB}$ to jointly approach the ideal radar and communication beamformers. Recalling that $L \le N_r \le N_t$, the communication channel matrix $\mathbf{H}$ has a rank of $L$, which supports a maximum  of $L$ independent data streams to be transmitted simultaneously.  Nevertheless, we use a digital beamformer with larger size $\mathbf{F}_{BB} \in \mathbb{C}^{K \times K}$ since we have to formulate extra beams towards the radar targets. In addition, the proposed method requires $\mathbf{F}_{BB}$ to have a full rank of $K$. Accordingly, \emph{K} RF chains are activated, leading to an ${N_t \times K}$ analog beamformer. The signal vector received at the UE can therefore be expressed as
\begin{equation}\label{eq39}
  {{\mathbf{y}}_{DL}} = {\mathbf{B}}\left( \Phi  \right)\operatorname{diag} \left( \bm{\beta}  \right){{\mathbf{A}}^T}\left( {{\Theta _1}} \right){{\mathbf{F}}_{RF}}{{\mathbf{F}}_{BB}}{\mathbf{s}} + {\mathbf{n}}_{DL},
\end{equation}
where ${\mathbf{n}}_{DL}$ denotes the noise vector with variance $\sigma_{DL}^2$, $\mathbf{s} \in \mathbb{C}^{K \times 1}$ denotes the transmit signal vector, which can be further decomposed as
\begin{equation}\label{eq40}
  {\mathbf{s}} = \left[ \begin{gathered}
  {{\mathbf{s}}_1} \hfill \\
  {{\mathbf{s}}_2} \hfill \\
\end{gathered}  \right],
\end{equation}
where ${{\mathbf{s}}_1}\in \mathbb{C}^{L \times 1}$  and ${{\mathbf{s}}_2}\in \mathbb{C}^{\left(K-L\right) \times 1}$ are statistically independent of each other. Each entry of $\mathbf{s}$ is assumed to follow a standard Gaussian distribution. Note that while both ${{\mathbf{s}}_1}$ and ${{\mathbf{s}}_2}$ are exploited for radar target detection, only ${{\mathbf{s}}_1}$ is exploited for DL communication whereas ${{\mathbf{s}}_2}$ contains no useful information, as the communication channel only supports transmission of \emph{L} independent data streams .
\\\indent Note that a pseudo inverse is unobtainable for the channel $\mathbf{H}$ since neither $\mathbf{H}\mathbf{H}^H$ nor $\mathbf{H}^H\mathbf{H}$ is invertible. Therefore, both transmit and receive beamformings are required for equalizing the channel. By introducing ${\mathbf{\tilde H}} \triangleq \operatorname{diag} \left( {\bm\beta}  \right){{\mathbf{A}}^T}\left( {{\Theta _1}} \right)$, the channel $\mathbf{H}$ can be equivalently expressed as ${\mathbf{H}} = {\mathbf{B}}\left( \Phi  \right){\mathbf{\tilde H}}$. Noting that both ${\mathbf{\tilde H}}$ and ${\mathbf{B}}\left( \Phi  \right)$ have a full rank of $L$, and that they have been estimated at the BS and the UE respectively, we can formulate the corresponding zero-forcing (ZF) beamformers as
\begin{equation}\label{eq41}
\begin{gathered}
  {{\mathbf{F}}_{BS}} = {{{\mathbf{\tilde H}}}^H}{\left( {{\mathbf{\tilde H}}{{{\mathbf{\tilde H}}}^H}} \right)^{ - 1}}, \hfill \\
  {{\mathbf{W}}_{UE}} = {\left( {{{\mathbf{B}}^H}\left( \Phi  \right){\mathbf{B}}\left( \Phi  \right)} \right)^{ - 1}}{{\mathbf{B}}^H}\left( \Phi  \right). \hfill \\
\end{gathered}
\end{equation}
While ${{\mathbf{W}}_{UE}}$ can be implemented as a fully-digital beamformer at the UE, ${{\mathbf{F}}_{BS}}$ can only be approximately approached by the hybrid array at the BS. In the meantime, the beamformer ${{\mathbf{F}}_{D}} = \mathbf{F}_{RF}\mathbf{F}_{BB}$ designed should also steer the beams towards all the \emph{K} targets. Note that this is equivalent to designing the covariance matrix of the transmit signal, which is formulated as
\begin{equation}\label{eq42}
\begin{gathered}
  {{\mathbf{R}}_s} = \mathbb{E}\left( {{{\mathbf{F}}_D}{\mathbf{s}}{{\mathbf{s}}^H}{\mathbf{F}}_D^H} \right) = {{\mathbf{F}}_D}\mathbb{E}\left( {{\mathbf{s}}{{\mathbf{s}}^H}} \right){\mathbf{F}}_D^H \hfill \\
   = {{\mathbf{F}}_{RF}}{{\mathbf{F}}_{BB}}{\mathbf{F}}_{BB}^H{\mathbf{F}}_{RF}^H. \hfill \\
\end{gathered}
\end{equation}
In what follows, we propose a low-complexity approach to the design of both ${{\mathbf{F}}_{RF}}$ and ${{\mathbf{F}}_{BB}}$.

\subsection{Low-complexity Approach for DFRC Hybrid Beamforming Design}
Based on the discussions above, a straightforward approach is to formulate each column of ${{\mathbf{F}}_{RF}}$ based on the steering vector associated with all the \emph{K} angles, yielding
\begin{equation}\label{eq43}
  {{\mathbf{F}}_{RF}}\left( {:,i} \right) = {{\mathbf{a}}^*}\left( {{\theta _i}} \right),\forall i.
\end{equation}
Nevertheless, the above ${{\mathbf{F}}_{RF}}$ does not guarantee having a desired transmit beampattern, which also depends on ${{\mathbf{F}}_{BB}}$. From (\ref{eq8}) and (\ref{eq42}), it becomes plausible that the transmit beampattern is solely dependent on ${{\mathbf{F}}_{RF}}$ if ${{\mathbf{F}}_{BB}}$ is an unitary matrix. To this end, we consider the following optimization problem by fixing ${{\mathbf{F}}_{RF}}$ as (\ref{eq43}), yielding
\begin{equation}\label{eq44}
\begin{gathered}
  \mathop {\min }\limits_{{{\mathbf{F}}_{BB}}} \;\left\| {{{\mathbf{F}}_{RF}}{{\mathbf{F}}_{BB}} - \left[ {{{\mathbf{F}}_{BS}},{{\mathbf{F}}_{aux}}} \right]} \right\|_F^2 \hfill \\
  s.t.\;\;\;{{\mathbf{F}}_{BB}}{\mathbf{F}}_{BB}^H = \frac{P}{{K{N_t}}}{{\mathbf{I}}_K}, \hfill \\
\end{gathered}
\end{equation}
where ${{\mathbf{F}}_{BS}} \in \mathbb{C}^{N_t \times L}$ is defined in (\ref{eq41}), and ${{\mathbf{F}}_{aux}} \in \mathbb{C}^{N_t \times \left({K-L}\right)} $ is an auxiliary matrix that is to be designed later. The scaling factor $ \frac{P}{{K{N_t}}}$ ensures satisfying the total transmit power budget of $\left\| {{{\mathbf{F}}_{RF}}{{\mathbf{F}}_{BB}}} \right\|_F^2 = P$. To be specific, the problem (\ref{eq44}) aims at approximating the fully-digital ZF beamformer ${{\mathbf{F}}_{BS}}$ by using the first $L$ columns of $\mathbf{F}_D$ while keeping the orthogonality of ${{{\mathbf{F}}_{BB}}}$.
\\\indent We can readily solve problem (\ref{eq44}) by obtaining its global optimum despite the non-convex constraint in $\mathbf{F}_{BB}$. Based on \cite{viklands2008algorithms,8386661}, problem (\ref{eq44}) can be classified as an orthogonal Procrustes problem (OPP), whose optimal solution can be formulated in closed-form as
\begin{equation}\label{eq45}
  {{\mathbf{F}}_{BB}}=\sqrt {\frac{P}{{K{N_t}}}} {\mathbf{\tilde U}}{{\mathbf{\tilde V}}^H},
\end{equation}
where
\begin{equation}\label{eq46}
  {\mathbf{{\tilde U}{\tilde \Sigma} }}{{\mathbf{{\tilde V}}}^H} = {\mathbf{F}}_{RF}^H\left[ {{{\mathbf{F}}_{BS}},{{\mathbf{F}}_{aux}}} \right]
\end{equation}
is the singular value decomposition (SVD) of ${\mathbf{F}}_{RF}^H\left[ {{{\mathbf{F}}_{BS}},{{\mathbf{F}}_{aux}}} \right]$.
\subsection{Spectral Efficiency Evaluation}
It can be noted that the above design is capable of guaranteeing the formulation of \emph{K} narrow beams towards radar targets. To show this, we write the transmit beampattern as
\begin{equation}\label{eq47}
\begin{gathered}
  d\left( \theta  \right) = \frac{P}{{K{N_t}}}{{\mathbf{a}}^T}\left( \theta  \right){{\mathbf{F}}_{RF}}{\mathbf{F}}_{RF}^H{{\mathbf{a}}^*}\left( \theta  \right) \hfill \\
   = \left\{ \begin{gathered}
  \frac{P}{{K{N_t}}}\left( {N_t^2 + \sum\limits_{\begin{subarray}{l}
  k = 1 \\
  k \ne i
\end{subarray}} ^K {{{\left| {{{\mathbf{a}}^T}\left( {{\theta}} \right){{\mathbf{a}}^*}\left( {{\theta _k}} \right)} \right|}^2}} } \right),\theta  = {\theta _i} \in \Theta , \forall i, \hfill \\
  \frac{P}{{K{N_t}}}\sum\limits_{k = 1}^K {{{\left| {{{\mathbf{a}}^T}\left( {{\theta}} \right){{\mathbf{a}}^*}\left( {{\theta _k}} \right)} \right|}^2}} ,\theta  \notin \Theta . \hfill \\
\end{gathered}  \right. \hfill \\
\end{gathered}
\end{equation}
When $N_t$ is sufficient large, ${{{\left| {{{\mathbf{a}}^T}\left( {{\theta _i}} \right){{\mathbf{a}}^*}\left( {{\theta _k}} \right)} \right|}^2}}$ will be much smaller than $N_t^2$ for any $i \ne k$, and thus a peak only appears if $\theta \in \Theta$.
\\\indent We then evaluate the performance of the communication by computing the spectral efficiency (SE). Let us firstly split the designed beamforming matrix as
\begin{equation}\label{eq48}
  {{\mathbf{F}}_D} = {{\mathbf{F}}_{RF}}{{\mathbf{F}}_{BB}}  = {{\mathbf{F}}_{RF}}\left[ {{{\mathbf{F}}_{BB,1}},{{\mathbf{F}}_{BB,2}}} \right],
\end{equation}
where ${{\mathbf{F}}_{BB,1}} \in {\mathbb{C}^{K \times L}},{{\mathbf{F}}_{BB,2}} \in {\mathbb{C}^{K \times \left( {K - L} \right)}}.$
By recalling (\ref{eq40}), and multiplying (\ref{eq39}) with ${{\mathbf{W}}_{UE}}$, the post-processing signal vector at the UE can be formulated by
\begin{equation}\label{eq49}
\begin{gathered}
  {{{\mathbf{\tilde y}}}_{DL}}
   = \sqrt \rho  {{\mathbf{W}}_{UE}}{\mathbf{H}}{{\mathbf{F}}_{RF}}{{\mathbf{F}}_{BB}}{\mathbf{s}} + {{\mathbf{W}}_{UE}}{\mathbf{n}} \hfill \\
   = \underbrace {\sqrt \rho  {{\mathbf{W}}_{UE}}{\mathbf{H}}{{\mathbf{F}}_{RF}}{{\mathbf{F}}_{BB,1}}{{\mathbf{s}}_1}}_\text{Useful Signal} + \underbrace {\sqrt \rho  {{\mathbf{W}}_{UE}}{\mathbf{H}}{{\mathbf{F}}_{RF}}{{\mathbf{F}}_{BB,2}}{{\mathbf{s}}_2}}_\text{Interference} \hfill \\
  \;\;\;\; +  {{\mathbf{W}}_{UE}}{\mathbf{n}}_{DL}. \hfill \\
\end{gathered}
\end{equation}
where $\rho$ stands for the average received power. The second term of (\ref{eq49}) is the interference imposed on the UE as it contains no useful information. The spectral efficiency is therefore given as
\begin{equation}\label{eq50}
R_{DL} = \log \det \left( \begin{gathered}
  {{\mathbf{I}}_L} + \frac{\rho }{L}{\mathbf{R}}_{in}^{ - 1}{{\mathbf{W}}_{UE}}{\mathbf{H}}{{\mathbf{F}}_{RF}}{{\mathbf{F}}_{BB,1}} \hfill \\
  \;\;\;\;\;\;\; \times {\mathbf{F}}_{BB,1}^H{\mathbf{F}}_{RF}^H{{\mathbf{H}}^H}{\mathbf{W}}_{UE}^H \hfill \\
\end{gathered}  \right),
\end{equation}
where ${\mathbf{R}}_{in}$ is the covariance matrix of the interference plus noise, which is
\begin{equation}\label{eq51}
\begin{gathered}
  {{\mathbf{R}}_{in}} = \rho {{\mathbf{W}}_{UE}}{\mathbf{H}}{{\mathbf{F}}_{RF}}{{\mathbf{F}}_{BB,2}}{\mathbf{F}}_{BB,2}^H{\mathbf{F}}_{RF}^H{{\mathbf{H}}^H}{\mathbf{W}}_{UE}^H \hfill \\
  \;\;\;\;\;\;\;\;\;\;\;+ \sigma _c^2{{\mathbf{W}}_{UE}}{\mathbf{W}}_{UE}^H. \hfill \\
\end{gathered}
\end{equation}
\subsection{Interference Reduction}
The enhancement of SE requires addressing the interference term in (\ref{eq49}). It can be observed that the interference power is mainly determined by ${{\mathbf{F}}_{RF}}{{\mathbf{F}}_{BB,2}}$, which is designed to approach $\mathbf{F}_{aux}$ in the optimization problem (\ref{eq44}). Hence, the choice of $\mathbf{F}_{aux}$ is key to the hybrid beamforming design.
\\\indent {\emph{HBF-Null Design:}} As an intuitive method, one may choose $\mathbf{F}_{aux}$ as a null-space projection (NSP) matrix, such that $\mathbf{\tilde H}\mathbf{F}_{aux} = \mathbf{0}$. This can be realized by firstly performing the SVD of $\mathbf{\tilde H}$, and then choosing the right singular vectors associated with zero singular values as the columns of $\mathbf{F}_{aux}$. By doing so, the solution of (\ref{eq44}) will satisfy that $\mathbf{\tilde H}{{\mathbf{F}}_{RF}}{{\mathbf{F}}_{BB,2}} \approx \mathbf{0}$ and thus $\mathbf{H}{{\mathbf{F}}_{RF}}{{\mathbf{F}}_{BB,2}} \approx \mathbf{0}$.
\\\indent {\emph{HBF-Opt Design:}} To further mitigate the interference, we consider another option by letting $\mathbf{F}_{aux} = \mathbf{0}$. While it is impossible to approach zero by multiplying the right side of ${\mathbf{F}}_{RF}$ with any unitary matrix, we show that such a method brings significant benefits by proving the following proposition.
\begin{proposition}
The interference can be completely eliminated by solving (\ref{eq44}) upon letting $\mathbf{F}_{aux} = \mathbf{0}$.
\end{proposition}
\renewcommand{\qedsymbol}{$\blacksquare$}
\begin{proof}
See Appendix.
\end{proof}
The intuition behind the HBF-Opt method is simple. Based on (\ref{eq45}) and (\ref{eq48}), ${\mathbf{F}}_{BB,1}$ is obtained by letting all the non-zero singular values of ${\mathbf{F}}_{RF}^H{{\mathbf{F}}_{BS}}$ be 1. As a result, ${\mathbf{F}}_{RF}{\mathbf{F}}_{BB,1}$ is an approximation of ${{\mathbf{F}}_{BS}}$. By letting $\mathbf{F}_{aux} = \mathbf{0}$, ${\mathbf{F}}_{BB,2}$ belongs to the null-space of ${\mathbf{F}}_{RF}^H{{\mathbf{F}}_{BS}}$, and thus belongs to the null-space of ${\mathbf{H}}{{\mathbf{F}}_{RF}}$ given the pseudo-inverse structure of ${{\mathbf{F}}_{BS}}$. Therefore, the interference of $\mathbf{s}_2$ is zero-forced.

\section{Stage 3: Radar Target Tracking and Uplink Communication}
After the joint transmission of radar and communication signals, the BS receives both the echo wave from the targets and the communication data from the UE. In this section, we propose a novel approach for joint radar target tracking and UL communication by relying on the knowledge of the previously estimated channel and target parameters.
\begin{figure}
    \centering
    \includegraphics[width=0.9\columnwidth]{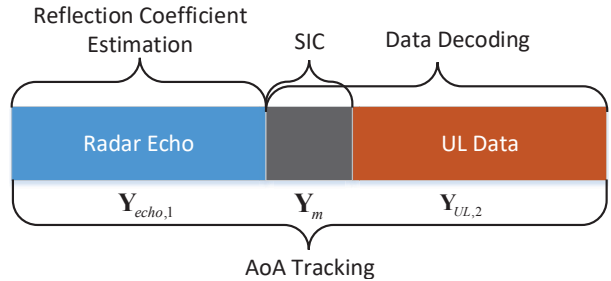}
    \caption{Overlapped receive signal model.}
    \label{fig:3}
\end{figure}
\subsection{Receive Signal Model}
According to the frame structure designed in Fig. 2, the signal received at the BS may fall into 2 categories: 1) Non-overlapped radar echo and UL communication signal and 2) overlapped signals. Since in the non-overlapped case both signals are interference-free, they can be readily processed using the conventional approaches. We therefore focus our attention on the overlapped case, where the radar and communication signals are partially interfering with each other.
\\\indent We show a generic model of the overlapped case in Fig. 3, where the overlapped period is marked as black. The received signal can be expressed as
\begin{equation}\label{eq59}
  {{\mathbf{Y}}_0} = \left[ {{{\mathbf{Y}}_{echo,1}},{{\mathbf{Y}}_m},{{\mathbf{Y}}_{UL,2}}} \right] \in {\mathbb{C}^{{N_t} \times {T_0}}},
\end{equation}
where ${{\mathbf{Y}}_{echo,1}} \in {\mathbb{C}^{{N_t} \times {\left(T-{\Delta T}\right)}}}$ denotes the non-interfered part of the radar echo wave, ${{\mathbf{Y}}_m} \in {\mathbb{C}^{{N_t} \times {{\Delta T}}}}$ represents the mixture of the echo wave and the communication signal received from the UE with ${\Delta T}$ being the length of the overlapping period, and finally ${{\mathbf{Y}}_{UL,2}} \in {\mathbb{C}^{{N_t} \times {\left(T_{c}-{\Delta T}\right)}}}$ stands for the non-interfered part of the UE signal with $T_{c}$ being the length of the UL frame. It can be readily seen that $T_0 = T+T_{c}-{\Delta T}$.
\\\indent By using the same notations from the previous sections, the above three signal matrices can be expressed as
\begin{equation}\label{eq60}
  {{\mathbf{Y}}_{echo,1}} = {\mathbf{A}}\left( {{\hat\Theta}  + \Delta \Theta } \right)\operatorname{diag} \left( {{\bm{\tilde \alpha }}} \right){{\mathbf{A}}^T}\left( {{\hat\Theta}  + \Delta \Theta } \right){{\mathbf{X}}_{r,1}} + {{\mathbf{Z}}_1},
\end{equation}
\begin{equation}\label{eq61}
\begin{gathered}
  {{\mathbf{Y}}_m} = {{\mathbf{Y}}_{echo,2}} + {{\mathbf{Y}}_{UL,1}}\hfill \\
   = {\mathbf{A}}\left( {{\hat\Theta}  + \Delta \Theta } \right)\operatorname{diag} \left( {{\bm{\tilde \alpha }}} \right){{\mathbf{A}}^T}\left( {{\hat\Theta}  + \Delta \Theta } \right){{\mathbf{X}}_{r,2}} \hfill \\
   + {\mathbf{A}}\left( {{{\hat\Theta} _1} + \Delta {\Theta _1}} \right)\operatorname{diag} \left( {{\bm{\tilde \beta }}} \right){{\mathbf{B}}^T}\left( {{\hat\Phi}  + \Delta \Phi } \right){{\mathbf{X}}_{UL,1}} + {{\mathbf{Z}}_m}, \hfill \\
\end{gathered}
\end{equation}
\begin{equation}\label{eq62}
\begin{gathered}
  {{\mathbf{Y}}_{UL,2}} = {\mathbf{A}}\left( {{{\hat\Theta} _1} + \Delta {\Theta _1}} \right)\operatorname{diag} \left( {{\bm{\tilde \beta }}} \right){{\mathbf{B}}^T}\left( {{\hat\Phi}  + \Delta \Phi } \right){{\mathbf{X}}_{UL,2}} \hfill \\
  \;\;\;\;\;\;\; + {{\mathbf{Z}}_2}, \hfill \\
\end{gathered}
\end{equation}
where ${\hat\Theta}$, ${\hat\Theta}_1 \subseteq {\hat\Theta}$ and $\hat\Phi$ contain the AoAs of all the \emph{K} targets, the AoAs and the AoDs of the UL channel (which are the AoDs and the AoAs of the DL channel, respectively) estimated in the last PRI, $\Delta \Theta$, $\Delta \Theta_1 $ and $\Delta \Phi$ represent accordingly the variations in these angles in the current PRI. Furthermore, $\bm {\tilde \alpha}$ contains the complex reflection coefficients of all the \emph{K} targets, while $\bm {\tilde \beta}$ contains the complex scattering coefficients of \emph{L} communication paths. Referring to (\ref{eq60}) and (\ref{eq62}), $\mathbf{X}_{r,1}$ and $\mathbf{X}_{UL,2}$ are the non-interfered parts of the radar and communication signals, while $\mathbf{X}_{r,2}$ and $\mathbf{X}_{UL,1}$ are the signals in the overlapped period, and finally $\mathbf{Z}_1, \mathbf{Z}_m, \mathbf{Z}_2$ denote the Gaussian noise matrices.
\\\indent To track the targets, the current AoAs and AoDs have to be estimated based on the previously estimated angles. As the angles are slowly varying as compared to the movement of the targets, we assume that these variations are relatively small. In contrast to the AoAs and AoDs, we assume that both $\bm {\tilde \alpha}$ and $\bm {\tilde \beta}$ are random realizations that are independent of those of the last PRI, and hence have to be estimated again. Furthermore, we denote the transmitted signal in the PRI as
\begin{equation}\label{eq63}
  {{\mathbf{X}}_r} = \left[ {{{\mathbf{X}}_{r,1}},{{\mathbf{X}}_{r,2}}} \right] \in {\mathbb{C}^{{N_t} \times {T} }},
\end{equation}
which has been precoded by $\mathbf{F}_{RF}$ and $\mathbf{F}_{BB}$ designed in Stage 2, where $\mathbf{F}_{RF}$ generates \emph{K} beams towards the estimated AoAs in ${\hat\Theta}$. Finally, the UL communication signal is given by
\begin{equation}\label{eq64}
  {{\mathbf{X}}_{UL}} = \left[ {{{\mathbf{X}}_{UL,1}},{{\mathbf{X}}_{UL,2}}} \right] \in {\mathbb{C}^{{N_r} \times T_c }},
\end{equation}
which has been precoded at the UE by $\mathbf{F}_{UE}$ in (\ref{eq34}) with the knowledge of the previously estimated $\hat \Phi$. Note that both ${{\mathbf{X}}_r}$ and ${{\mathbf{X}}_{UL}}$ are assumed to be Gaussian distributed following the previous assumptions. For the sake of convenience, we employ the assumption that the BS can reliably identify the beginning of ${{\mathbf{X}}_{UL}}$. This can be realized by inserting synchronization sequences at the beginning of the ${{\mathbf{X}}_{UL}}$. The designed sequences should be orthogonal to the radar signal ${{\mathbf{X}}_r}$, such that the interference of the echo wave can be mitigated at the synchronization stage\footnote{Note that such synchronization sequences can be easily formulated as the null-space projection matrix of the radar signal. Nevertheless, the data sequences that contain information from the UE are unlikely to be orthogonal to the radar signal. Hence, we still need to mitigate the radar interference when processing the communication signal after synchronization.}.
\\\indent In what follows, we propose approaches for both target tracking and UL signal processing.
\subsection{Target Tracking}
After receiving $\mathbf{Y}_0$, the first step is analog combination, which gives us
\begin{equation}\label{eq65}
\begin{gathered}
  {{{\mathbf{\tilde Y}}}_0} = {{\mathbf{W}}_{RF}}{{\mathbf{Y}}_0} \hfill \\
   = \left[ {{{\mathbf{W}}_{RF}}{{\mathbf{Y}}_{echo,1}},{{\mathbf{W}}_{RF}}{{\mathbf{Y}}_m},{{\mathbf{W}}_{RF}}{{\mathbf{Y}}_{UL,2}}} \right] \in {\mathbb{C}^{{N_{RF}} \times {T_0}}}, \hfill \\
\end{gathered}
\end{equation}
where we activate all $N_{RF}$ RF chains to formulate an analog combination matrix ${{\mathbf{W}}_{RF}} \in {\mathbb{C}^{{N_{RF}} \times {N_t}}}$. To exploit the knowledge of the estimated angles in $\hat \Theta$, the first \emph{K} rows of ${{\mathbf{W}}_{RF}}$ (which represent the phase shifters linked with the first \emph{K} RF chains) are set as
\begin{equation}\label{eq66}
  {{\mathbf{W}}_{RF}}\left( {k,:} \right) = {{\mathbf{a}}^H}\left( {{{\hat \theta }_k}} \right),\forall k,
\end{equation}
which indicates that the receive beams are pointing to the previously estimated AoAs. The phase shifters in the remaining RF chains are randomly set, thus for creating redundant observations of the received data in order to improve the estimation accuracy.
\\\indent An important fact that can be observed from (\ref{eq60})-(\ref{eq62}) is that the mutual interference signal in (\ref{eq61}) will not degrade the AoA estimation performance. Instead, it may provide benefits in estimating some of the AoAs. This is because the BS receives both the echo waves and the communication signals from the angles in $\Theta_1 + \Delta \Theta_1$. As a result, the signal associated with these angles may have higher power than that associated with others, hence leading to better estimation performance.
\\\indent Given the small variations in the AoAs, one may search in the small intervals within each ${\hat \theta}_k,\forall k$ instead of searching the whole angular domain. We therefore propose to apply the MUSIC algorithm to ${{{\mathbf{\tilde Y}}}_0}$ for estimating the AoAs. For each ${\hat \theta}_k$, we search for peaks in the MUSIC spectrum (\ref{eq26}) within $\left[ {{{\hat \theta }_k} - {\Delta _{\max }},{{\hat \theta }_k} + {\Delta _{\max }}} \right]$, where $\Delta _{\max }$ is the maximum angular variation of the targets.
\subsection{Uplink Communication}
In this subsection, we propose a promising technique for estimating the remaining target parameters and decode the communication signals. Since ${{\mathbf{Y}}_{echo,1}}$ is not interfered by the communication signal, it can be used to estimate the target reflection coefficients $\bm{\tilde \alpha}$. With the estimated AoAs at hand, one can apply the APES approach to obtain an estimate of ${\tilde{\alpha}}_k$, i.e., ${\hat{\tilde \alpha}}_k$ for each angle.
\\\indent The communication signal can then be recovered by the SIC approach. Given the estimated parameters and ${{\mathbf{X}}_r}$, the target reflections can be reconstructed as
\begin{equation}\label{eq67}
{{{\mathbf{\hat Y}}}_{echo}} = {\mathbf{A}}\left( {\hat \Theta  + \Delta \hat \Theta } \right)\operatorname{diag} \left( {\hat {\tilde {\bm\alpha}} } \right){{\mathbf{A}}^T}\left( {\hat \Theta  + \Delta \hat \Theta } \right){{\mathbf{X}}_r},
\end{equation}
where $\Delta \hat \Theta$ denotes the estimated variations of AoAs. Note that by multiplying ${\mathbf{W}}_{RF}$, the $N_t \times T_0$ matrix ${{\mathbf{Y}}_0}$ has been mapped to a lower-dimensional space having the size of $N_{RF} \times T_0$. Therefore, one can only recover the communication signal after low-complexity analog combination. By subtracting the radar signal estimated, the interfered communication signal in $\mathbf{Y}_m$ can be estimated as
\begin{equation}\label{eq68}
\begin{gathered}
  {{\mathbf{W}}_{RF}}{{{\mathbf{\hat Y}}}_{UL,1}} = {{\mathbf{W}}_{RF}}{{\mathbf{Y}}_m} \hfill \\
   - {{\mathbf{W}}_{RF}}{\mathbf{A}}\left( {\hat \Theta  + \Delta \hat \Theta } \right)\operatorname{diag} \left( {\hat {\tilde {\bm\alpha}} }\right){{\mathbf{A}}^T}\left( {\hat \Theta  + \Delta \hat \Theta } \right){{\mathbf{X}}_{r,2}}. \hfill \\
\end{gathered}
\end{equation}
Based on the above, the whole UL signal after analog combination can be expressed as
\begin{equation}\label{eq69}
  {{\mathbf{W}}_{RF}}{{{\mathbf{\hat Y}}}_{UL}} = \left[ {{{\mathbf{W}}_{RF}}{{{\mathbf{\hat Y}}}_{UL,1}},{{\mathbf{W}}_{RF}}{{\mathbf{Y}}_{UL,2}}} \right].
\end{equation}
Since the UL signal has been precoded by (\ref{eq34}) at the UE, the steering matrix ${{\mathbf{B}}^T}\left( {{\hat\Phi}  + \Delta \Phi } \right)$ has been eliminated with limited errors. The BS can simply obtain the estimates of the path-losses $\hat{\tilde {\bm\beta}}$ by the LS approach with the help of the known synchronization sequence, and construct a baseband ZF beamformer by computing the following pseudo-inverse
\begin{equation}\label{eq70}
  {{\mathbf{W}}_{BB}} = {\left( {{{\mathbf{W}}_{RF}}{\mathbf{A}}\left( {{{\hat \Theta }_1} + \Delta {{\hat \Theta }_1}} \right)\operatorname{diag} \left( \hat{\tilde {\bm\beta}} \right)} \right)^\dag }.
\end{equation}
Upon multiplying ${{\mathbf{W}}_{RF}}{{{\mathbf{\hat Y}}}_{UL}}$ by ${{\mathbf{W}}_{BB}}$, the communication symbols can be finally decoded. For clarity, we summarize the signal processing procedures of Stage 3 in Algorithm 2.
\begin{algorithm}
\caption{Stage 3: Radar Target Tracking and UL Communication}
\label{alg:B}
\begin{algorithmic}
    \STATE \textbf{Step 1}: BS receives both target echoes and UL signals that are partially overlapped with each other.
    \STATE \textbf{Step 2}: BS formulates an analog combiner ${{\mathbf{W}}_{RF}}$ based on estimated $\hat \Theta$ in the last PRI.
    \STATE \textbf{Step 3}: BS estimates the reflection coefficients and the angular variation $\Delta \Theta$ by searching in a small interval within each ${\hat \theta_k} \in {\hat \Theta}$.
    \STATE \textbf{Step 4}: BS recovers the radar echoes based on the estimates from Step 3, and removes the radar interference in the overlapped part of the received signal.
    \STATE \textbf{Step 5}: BS formulates a ZF beamformer to equalize the communication channel, and decodes the UL data.
\end{algorithmic}
\end{algorithm}
\subsection{Spectral Efficiency Evaluation}
We round off this section by proposing a performance metric for the UL communication. While the estimated radar interference has been subtracted from ${{\mathbf{Y}}_m}$, there will still be some residual interference potentially degrading the communication performance. The residual interference can be expressed as
\begin{equation}\label{eq71}
\begin{gathered}
  {{\mathbf{Y}}_{res}} = {\mathbf{A}}\left( {\hat \Theta  + \Delta \Theta } \right)\operatorname{diag} \left( \bm{\tilde \alpha } \right){{\mathbf{A}}^T}\left( {\hat \Theta  + \Delta \Theta } \right){{\mathbf{X}}_{r,2}} \hfill \\
   - {\mathbf{A}}\left( {\hat \Theta  + \Delta \hat \Theta } \right)\operatorname{diag} \left( {\hat {{\tilde {\bm\alpha}} }} \right){{\mathbf{A}}^T}\left( {\hat \Theta  + \Delta \hat \Theta } \right){{\mathbf{X}}_{r,2}} \in \mathbb{C}^{N_t \times \Delta T}. \hfill \\
\end{gathered}
\end{equation}
Fortunately, the above interference will only be active during the first $\Delta T$ symbols, in which case the spectral efficiency can be given by
\begin{equation}\label{eq72}
  {R_1} = \log \det \left( \begin{gathered}
  {{\mathbf{I}}_L} + \frac{\rho }{L}{\mathbf{R}}_{in}^{ - 1}{{\mathbf{W}}_{BB}}{{\mathbf{W}}_{RF}}{\mathbf{H}}{{\mathbf{F}}_{UE}} \hfill \\
   \;\;\;\;\;\;\; \times {\mathbf{F}}_{UE}^H{{\mathbf{H}}^H}{\mathbf{W}}_{RF}^H{\mathbf{W}}_{BB}^H \hfill \\
\end{gathered}  \right),
\end{equation}
where
\begin{equation}\label{eq73}
  {{\mathbf{R}}_{in}} = {{\mathbf{W}}_{BB}}{{\mathbf{W}}_{RF}}\left(\frac{1}{{\Delta T}} {{{\mathbf{Y}}_{res}}{\mathbf{Y}}_{res}^H + \sigma _{UL}^2{{\mathbf{I}}_{{N_t}}}} \right){\mathbf{W}}_{RF}^H{\mathbf{W}}_{BB}^H
\end{equation}
is the covariance matrix of the interference plus noise, and $\rho$ is the average received power. During the interference-free period having a length of $T_c - \Delta T$, the spectral efficiency can be expressed as
\begin{equation}\label{eq74}
  {R_2} = \log \det \left( \begin{gathered}
  {{\mathbf{I}}_L} + \frac{\rho }{{L\sigma _{UL}^2}}{\left( {{\mathbf{W}}_{RF}^H{\mathbf{W}}_{BB}^H} \right)^\dag }{\mathbf{H}}{{\mathbf{F}}_{UE}} \hfill \\
   \;\;\;\;\;\;\; \times {\mathbf{F}}_{UE}^H{{\mathbf{H}}^H}{\mathbf{W}}_{RF}^H{\mathbf{W}}_{BB}^H \hfill \\
\end{gathered}  \right).
\end{equation}
The overall UL SE can be computed as the weighted summation of $R_1$ and $R_2$, which is
\begin{equation}\label{eq75}
  {R_{UL}} = \frac{{\Delta T}}{{{T_c}}}{R_1} + \frac{{{T_c} - \Delta T}}{{{T_c}}}{R_2}.
\end{equation}

\section{Numerical Results}
In this section, we provide numerical results to validate the performance of the proposed DFRC framework. Without loss of generality, the BS is assumed to be equipped with $N_t = 64$ antennas and $N_{RF} = 16$ RF chains, which communicates with a UE having $N_r = 10$ antennas. Unless otherwise specified, we assume that the BS is detecting $K = 8$ targets, wherein $L = 4$ of them act as the scatterers in the communication channel. Unless otherwise specified, all the AoAs and AoDs are randomly drawn from the interval of $\left[-90^\circ,90^\circ\right]$, which has been uniformly split into 180 slices. All the reflection and the scattering coefficients are assumed to obey the standard complex Gaussian distribution.
\begin{figure}
    \centering
    \includegraphics[width=0.95\columnwidth]{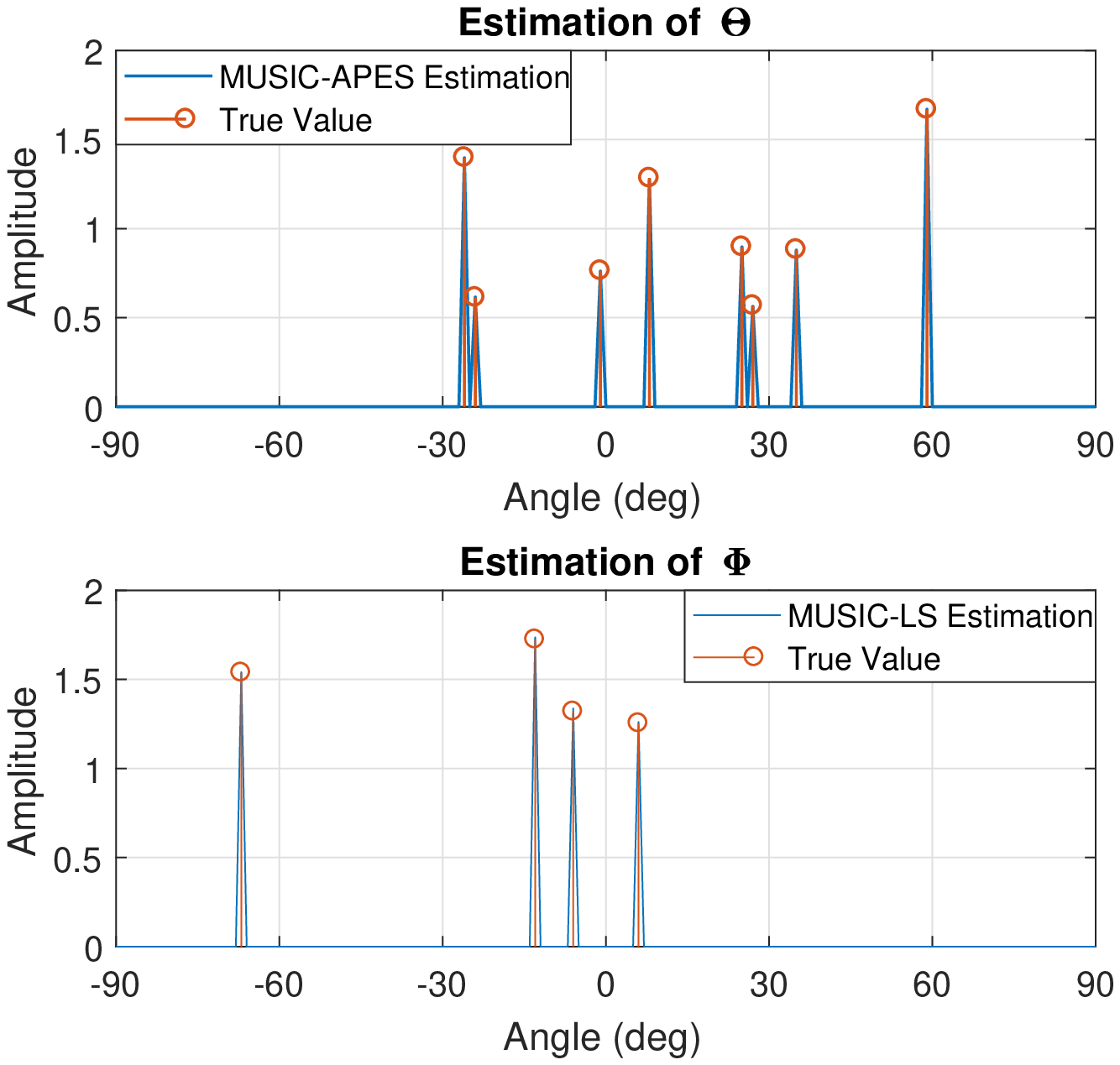}
    \caption{Angle estimation performance for Case 1 by using (27) and (31), $\text{SNR} = 10\text{dB}$, $T = 100$.}
    \label{fig:4}
\end{figure}
\begin{figure}
    \centering
    \includegraphics[width=0.95\columnwidth]{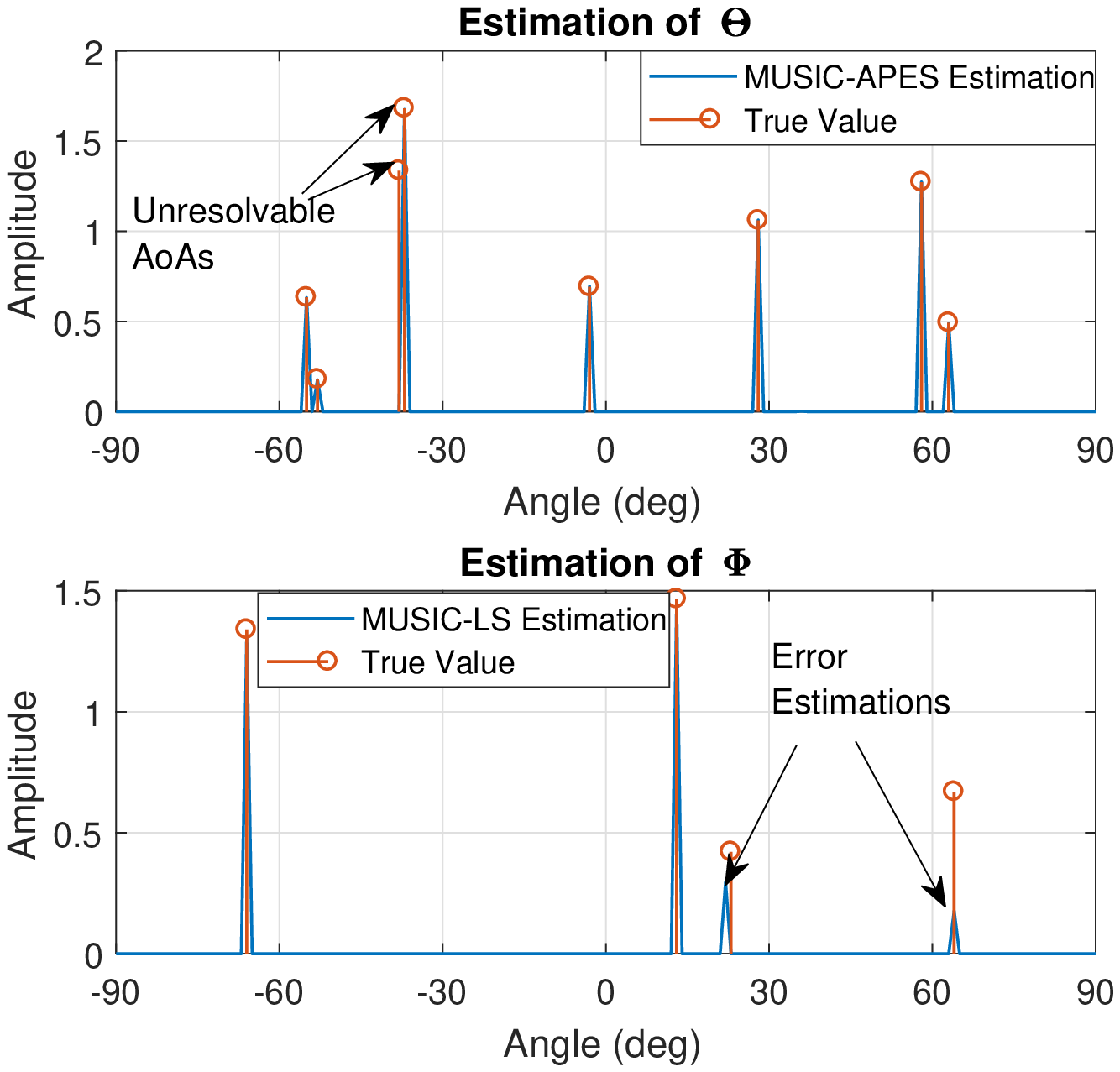}
    \caption{Angle estimation performance for Case 2 by using (27) and (31), $\text{SNR} = 0\text{dB}$, $T = 100$.}
    \label{fig:5}
\end{figure}
\begin{figure}
    \centering
    \includegraphics[width=0.95\columnwidth]{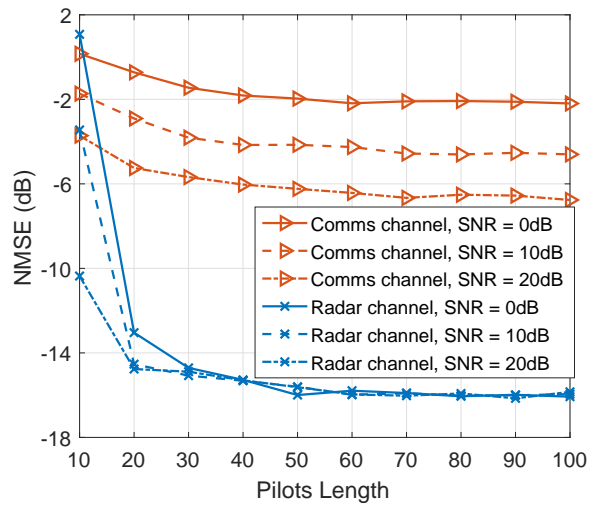}
    \caption{Estimation NMSE of both radar and communication channel.}
    \label{fig:6}
\end{figure}
\subsection{Radar Target Search and Channel Estimation}
We first show the performance of Stage 1 in Figs. 4-6 with the aid of target search and channel estimation results. More specifically, in Fig. 4, we show the target estimation performance for a single channel realization at $\text{SNR} = 10\text{dB}$ for both DL and UP. We use a $64 \times 100$ LFM signal matrix as the DP, and a $4 \times 4$ identity matrix as the UP. We compare the estimated results to the true values for both $\Theta$ and $\Phi$. It can be seen that the proposed MUSIC-APES and MUSIC-LS algorithms obtain accurate estimates of all the targets/scatterers at both the BS and the UE. It is worth highlighting that the MUSIC-APES has a superior angular resolution of $2^\circ$, which accurately differentiates the angle pairs $\left[-26^\circ,-24^\circ\right]$ and $\left[25^\circ,27^\circ\right]$.
\\\indent We then consider another example in Fig. 5 at $\text{SNR} = 0\text{dB}$, where there are two targets close to each other at the angles of $\left[-38^\circ,-37^\circ\right]$. In this case, the BS fails to identify the target at $-38^\circ$ despite that it successfully estimates all the other 7 targets. Furthermore, the UE makes a wrong estimation at the angle of $\phi_3 = 23^\circ$ with an error of $1^\circ$. As a result, the estimations of the scattering coefficients $\beta_3$ and $\beta_4$ show large errors compared to the true values. This suggests that when the targets are too close to each other, the accumulated angular estimation errors will have an impact on the estimation performance of the path coefficients. Nevertheless, since most of the angles are accurately estimated, the communication performance will only be marginally affected.
\\\indent In Fig. 6, we show the normalized mean-squared error (NMSE) of both the radar and communication channels upon varying the SNR and the DP length by averaging 8000 random channel realizations. It can be observed that the NMSE decreases in general with the growth of both parameters. Note that the estimation performance of the radar channel is far better than that of the communication channel. This is because the BS employs 64 antennas to estimate the AoAs, while the UE only has 10 antennas. In addition, the length of the UP has to be very short (in our case it is fixed as $L = 4$) given the frame structure we proposed in Fig. 2, which might lead to estimation errors in $\bm \beta$. We will show in the next subsection that fortunately the overall communication performance is good, despite the estimation errors in Stage 1.
\subsection{Radar Transmit Beamforming and Downlink Communication}
\begin{figure}
    \centering
    \includegraphics[width=0.95\columnwidth]{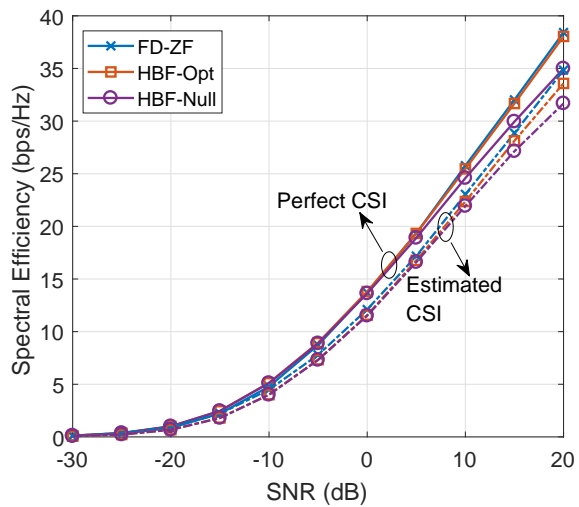}
    \caption{Spectral efficiency of the DL communication by using (47).}
    \label{fig:7}
\end{figure}
\begin{figure}
    \centering
    \includegraphics[width=0.95\columnwidth]{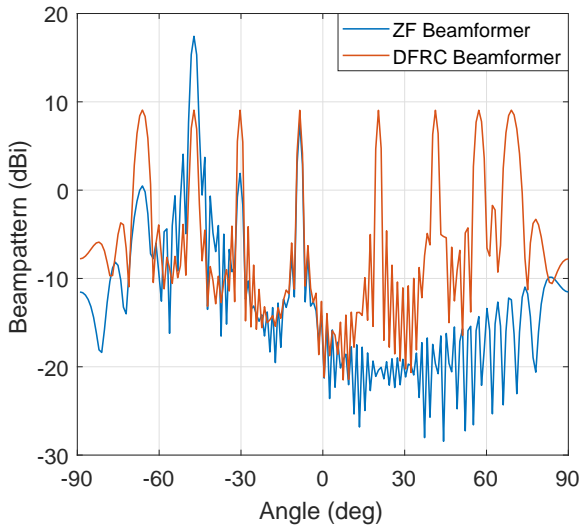}
    \caption{Transmit beampatterns for the communication-only ZF beamformer and the proposed DFRC beamformers.}
    \label{fig:8}
\end{figure}
Figs. 7-9 characterize the performance of Stage 2 in terms of the SE of the DL communication, the transmit beampattern and the number of the targets. In Fig. 7, we show the SE versus SNR of both perfect CSI and estimated CSI cases, where `FD-ZF' denotes fully digital ZF beamforming, `HBF-Opt' and `HBF-Null' represent the hybrid beamforming designs proposed in Sec. VI-D with $\mathbf{F}_{aux}$ being zero and NSP matrices, respectively. There are slight SE performance-losses for the cases with estimated CSI, which suggests that the proposed channel estimation method guarantees a satisfactory communication performance. Furthermore, we see that in both the perfect and estimated CSI cases, the HBF-Opt design outperforms the HBF-Null design by approaching the performance of the fully digital ZF beamformer, which verifies our derivation on interference reduction.
\\\indent Fig. 8 shows the transmit beampattern for both the communication-only ZF beamformer and for the HBF beamformers designed for the DFRC system proposed. While the HBF-Opt and the HBF-Null designs employ different unitary matrices as $\mathbf{F}_{BB}$, the resultant beampatterns are the same since they use the same $\mathbf{F}_{RF}$. It can be seen that the ZF beamformer only formulates beams towards 4 scatterers in the communication channel, and thus fails to track the extra 4 targets. By contrast, the proposed DFRC beamformer successfully generates 8 beams towards all the 8 targets.
\begin{figure}
    \centering
    \includegraphics[width=0.95\columnwidth]{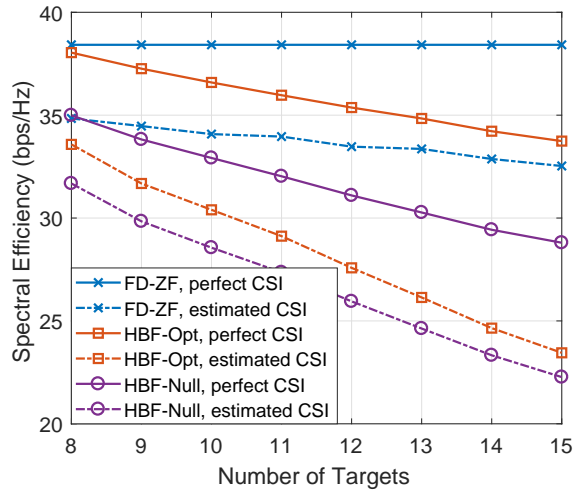}
    \caption{Tradeoff between the spectral efficiency and the number of targets, $\text{SNR} = 20\text{dB}$.}
    \label{fig:9}
\end{figure}
\\\indent To explicitly illustrate the performance tradeoff between radar and communication, we show in Fig. 9 the DL spectral efficiency by varying the number of targets at $\text{SNR} = 20\text{dB}$, where we fix the number of scatterers in the communication channel as $L = 4$, and increase the total number of targets from $K = 8$ to $15$. Since illuminating more targets requires more transmit power, less power is allocated to beams towards AoAs of the communication scatterers, leading to a reduced SINR. As a result, the DL SE decreases upon increasing the number of targets. Again, the SE of the HBF-Opt design is larger than that of the HBF-Null design. It is also interesting to observe the reduced SE of the fully digital ZF beamformer using estimated CSI, as the channel estimation becomes inaccurate owing to the newly added targets.
\subsection{Radar Target Tracking and Uplink Communication}
\begin{figure}
    \centering
    \includegraphics[width=0.95\columnwidth]{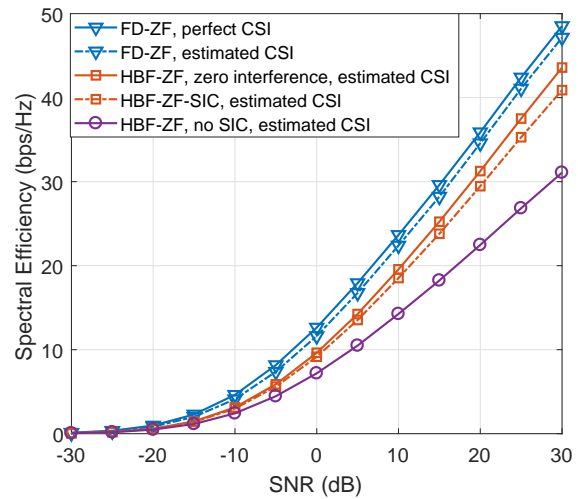}
    \caption{Spectral efficiency of the UL communication by using (65), 30\% overlapped ratio.}
    \label{fig:10}
\end{figure}
\begin{figure}
    \centering
    \includegraphics[width=0.95\columnwidth]{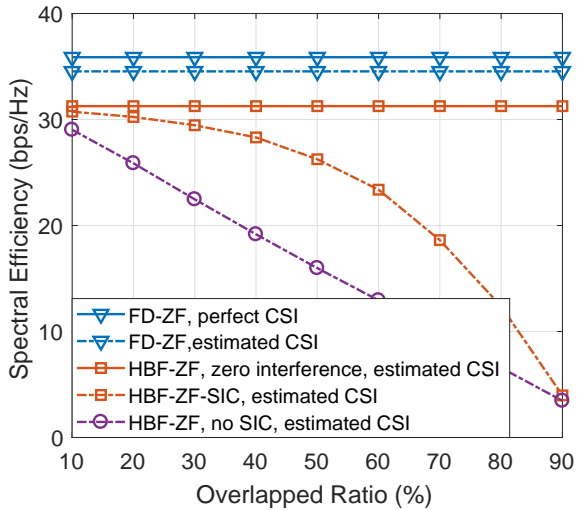}
    \caption{Tradeoff between the spectral efficiency and the overlapped ratio, $\text{SNR} = 20\text{dB}$.}
    \label{fig:11}
\end{figure}
\begin{figure}
    \centering
    \includegraphics[width=0.95\columnwidth]{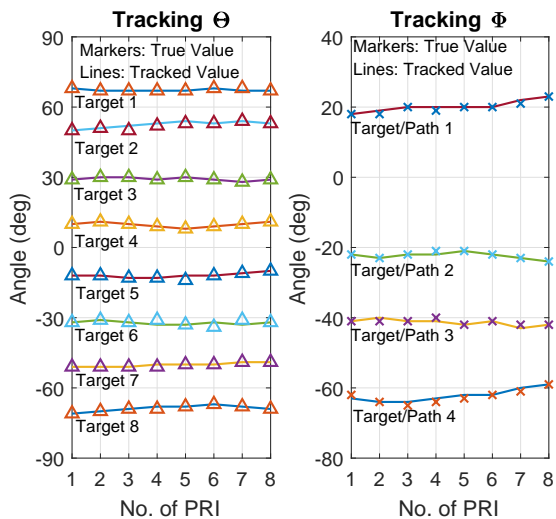}
    \caption{Angle tracking performance at the BS and the UE, $\text{SNR} = -20\text{dB}$.}
    \label{fig:12}
\end{figure}
Finally, we provide results for Stage 3 in Figs. 10-13, where we assume that the angle parameters of all the 8 targets of the previous PRI are perfectly known, based on which the DFRC system tracks the variation of the angles in the current PRI, while performing UL communications. As the angle parameters typically vary slowly in realistic scenarios, we assume without loss of generality that the variation of each angle is less than $\Delta_{\max} = 1^\circ$ at each PRI, which is reasonable for a PRI of a few of milliseconds. The DL and UL frame lengths are set as 140. The communication signal and the target echo wave are overlapped with each other, and share the same SNR. While it is known that the equivalent SNR scenario is the worst case for the SIC-based approaches, we will show next that our method can still achieve good performance.
\\\indent Fig. 10 shows the UL SE performance of the proposed approach in Sec. VII. It is noteworthy that by using the SIC method proposed, the SE of the communication significantly increases compared to the cases with full radar echo interference in the overlapped period. Fig. 11 further illustrates the UL SE performance given the increased overlapped period $\Delta T$, where the overlapping ratio is defined as $\Delta T/T$. We see that the SE becomes worse for longer overlapped period, in which case the interference of the radar echo is not cancelled thoroughly, and the residual interference power may have a grave impact on the UL communication performance. When the overlapped period is short, the performance gain obtained by the SIC approach is marginal since the interference from the radar echo is small enough. On the other hand, when the overlapped ratio is greater than 90\%, the BS fails to recover the radar signal, and thus is unable to cancel the interference by using the SIC, which also leads to modest performance gain. Nevertheless, in most overlapping cases, the SIC approach works well by considerably improving the SE.
\\\indent In Fig. 12, we evaluate the performance of the proposed target tracking approach, where we compare the tracking results and the true variation for the AoAs $\Theta$ of the targets and the AoDs $\Phi$ of the scattering paths at $\text{SNR} = -20\text{dB}$ for both target echoes and the communication signals. Note that the angles in $\Theta$ are estimated at the BS using both the target echoes and the UL signals, while the angles in $\Phi$ are estimated at the UE. It can be seen that all the angles can be accurately tracked with slight tracking errors despite the low SNR, which verifies again the effectiveness of the proposed method. Similar results are observed in Fig. 13, where we show the root-mean-squared-error (RMSE) of the proposed target tracking approach versus the SNR. It is shown that the RMSE for all the estimations is less than $1^\circ$ at most of the SNR values for both $\Delta_{\max} = 1^\circ$ and $2^\circ$.
\begin{figure}
    \centering
    \includegraphics[width=0.95\columnwidth]{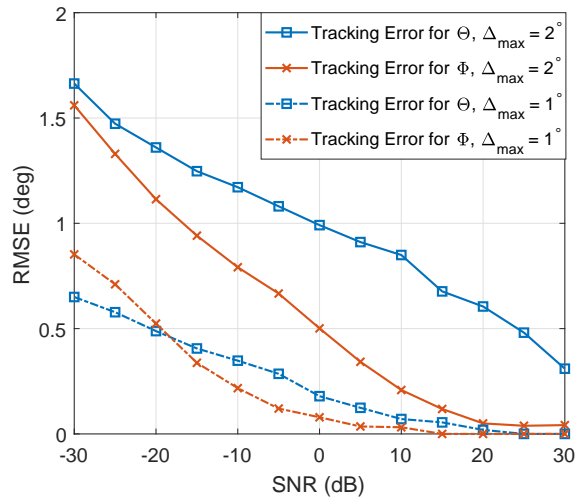}
    \caption{Angle tracking RMSE vs. SNR.}
    \label{fig:13}
\end{figure}

\section{Conclusion and Future Research}
\subsection{Summary of the Proposed Approaches}
In this paper, we have reviewed the application scenarios and recent research progress in the area of communication and radar spectrum sharing (CRSS). We have proposed a novel dual-functional radar-communication (DFRC) system architecture that operates in the mmWave band, and is equipped with a massive MIMO antenna array and a hybrid analog-digital beamforming structure. We have further designed a novel TDD frame structure that can unify the radar and communication operations into 3 stages, namely 1) radar target search and channel estimation, 2) radar transmit beamforming and DL communication and 3) radar target tracking and UL communication. Accordingly, we have proposed joint signal processing strategies for each stage. In Stage 1, we aim for estimating the communication channel and searching for potential targets using orthogonal LFM signals generated by the HAD structure, while identifying the communication paths from the radar targets. In Stage 2, we have designed both analog and digital precoders for generating directional beams towards all the targets and scatterers, while pre-equalizing the impact of the communication channel. Finally in Stage 3, we have proposed a joint scheme for tracking the angular variation of all the targets, while decoding the UL communication signals by using the SIC approach. Simulation results have been provided to validate the proposed approaches, showing the feasibility of realizing both radar and communication functionalities on a single mmWave BS.
\subsection{Future Works}
While a number of contributions have been made towards radar-communication coexistence and joint radar-communication systems, the topic remains to be further explored within a broader range of constraints and scenarios. To this end, we list in the following a number of future research directions in the area.
\\\indent \emph{1) Learning based CRSS}
\\\indent A key challenge for CRSS is to distinguish between the echoes from targets and communication signals from users in the presence of noise and interference. In addition to the proposed joint receiver design for the mmWave system considered, it is also viable to apply machine learning (ML) based approaches, such as the independent component analysis (ICA) algorithm, for signal classification in more generic scenarios, given the independent statistical characteristics of the two kinds of signals. A recent example can be found in \cite{8233171} where the compressed sensing (CS) approach is employed for joint parameter estimation and symbol demodulation. It is expected that by using advanced ML based techniques, the receiver design for CRSS can be well-addressed.
\\\indent \emph{2) Security issues}
\\\indent Recent CRSS research raised security and privacy concerns. By sharing the spectrum with communication systems, the military radar may unintentionally give away vital information to commercial users, or even worse, to the adversary eavesdroppers. To this end, physical layer security must be considered in the CRSS scenarios, where a possible method is that radar actively transmits artificial noise (AN) to the adversary target to contaminate the eavesdropping, while formulating desired beampatterns. In the meantime, the communication performance also has to be guaranteed. Accordingly, a number of performance trade-offs involving the radar detection and estimation performance, the communication rate and the secrecy rate remain to be studied. Some initial works on this topic can be found in \cite{8327462,CHALISE2018282,radar_privacy}.
\\\indent \emph{3) DFRC for V2X}
\\\indent As an important application scenario of the DFRC system, vehicular networks have recently drawn much attention from both industry and academia, where joint sensing and communications at the mmWave band is required. While the proposed approaches in this paper focus on mmWave cellular systems, it can be extended to V2X applications with the consideration of specific channel models for vehicle-to-vehicle (V2V) and vehicle-to-infrastructure (V2I) scenarios. Again, such schemes call for the design of novel beamforming/signaling approaches \cite{8057284,8114253,8642926}.
\\\indent \emph{4) Information theory aspects}
\\\indent To gain more in-depth insight into DFRC systems, information theoretical analysis is indispensable for revealing the fundamental performance limit. While existing contributions have considered the DFRC UL \cite{7279172} as well as coexisting radar and communication systems \cite{8070342}, the DL DFRC channel needs further investigations. Here the key point is to view the radar targets as virtual energy receivers, and hence the DFRC transmission can be seen as the allocation of information and energy resources in the NLoS and LoS channels. From a higher-level perspective, one can also view the radar target as a relay, which receives the probing waveform and forwards it back to the radar, with its own parameter information being embedded in the echo wave. As such, the target detection problem can be analyzed using the information theory of the relay channel, where a number of information metrics can be defined. It is believed that such analysis could help us to understand the intrinsic nature of the DFRC systems, and point us to the essential system design criteria.

\appendix[Proof of Proposition 1]
Let us denote the SVD of ${\mathbf{F}}_{RF}^H{{\mathbf{F}}_{BS}}$ as
\begin{equation}\label{eq52}
  {\mathbf{F}}_{RF}^H{{\mathbf{F}}_{BS}} = \left[ {{{{\mathbf{\tilde U}}}_s},{{{\mathbf{\tilde U}}}_n}} \right]\left[ \begin{gathered}
  {{{\mathbf{\tilde \Sigma }}}_s} \hfill \\
  {\mathbf{0}} \hfill \\
\end{gathered}  \right]{\mathbf{\tilde V}_s},
\end{equation}
where ${{{\mathbf{\tilde U}}}_s} \in {\mathbb{C}^{K \times L}}$ and ${{{\mathbf{\tilde V}}}_s} \in {\mathbb{C}^{L \times L}}$ contain the left and right singular vectors associated with non-zero singular values, and ${{{\mathbf{\tilde U}}}_n} \in {\mathbb{C}^{K \times \left(K-L\right)}}$ contains the left singular vectors corresponding to zero singular values. We then compute the optimal solution of (\ref{eq44}) when $\mathbf{F}_{aux} = \mathbf{0}$. Note that
\begin{equation}\label{eq53}
  {\mathbf{F}}_{RF}^H\left[ {{{\mathbf{F}}_{BS}},{\mathbf{F}_{aux}}} \right] = \left[ {{{{\mathbf{\tilde U}}}_s},{{{\mathbf{\tilde U}}}_n}} \right]\left[ {\begin{array}{*{20}{c}}
  {{{{\mathbf{\tilde \Sigma }}}_s}}&{} \\
  {}&{\mathbf{0}}
\end{array}} \right]\left[ {\begin{array}{*{20}{c}}
  {{{{\mathbf{\tilde V}}}_s}}&{} \\
  {}&{{{{\mathbf{\tilde V}}}_n}}
\end{array}} \right],
\end{equation}
which is the SVD of ${\mathbf{F}}_{RF}^H\left[ {{{\mathbf{F}}_{BS}},{\mathbf{F}_{aux}}} \right]$ for $\mathbf{F}_{aux} = \mathbf{0}$, where ${{{{\mathbf{\tilde V}}}_n}}$ is an arbitrary $(K-L) \times (K-L)$ unitary matrix. The optimal solution to problem (\ref{eq44}) can therefore be obtained in the form
\begin{equation}\label{eq54}
\begin{gathered}
  {{\mathbf{F}}_{BB}} = \sqrt {\frac{P}{{K{N_t}}}} \left[ {{{{\mathbf{\tilde U}}}_s},{{{\mathbf{\tilde U}}}_n}} \right]\left[ {\begin{array}{*{20}{c}}
  {{{{\mathbf{\tilde V}}}_s}}&{} \\
  {}&{{{{\mathbf{\tilde V}}}_n}}
\end{array}} \right] \hfill \\
   = \sqrt {\frac{P}{{K{N_t}}}} \left[ {{{{\mathbf{\tilde U}}}_s}{{{\mathbf{\tilde V}}}_s},{{{\mathbf{\tilde U}}}_n}{{{\mathbf{\tilde V}}}_n}} \right]. \hfill \\
\end{gathered}
\end{equation}
It follows that
\begin{equation}\label{eq55}
  {{\mathbf{F}}_{BB,1}} = \sqrt {\frac{P}{{K{N_t}}}} {{{\mathbf{\tilde U}}}_s}{{{\mathbf{\tilde V}}}_s},{{\mathbf{F}}_{BB,2}} = \sqrt {\frac{P}{{K{N_t}}}} {{{\mathbf{\tilde U}}}_n}{{{{\mathbf{\tilde V}}}_n}}.
\end{equation}
It can be readily verified that ${{\mathbf{F}}_{BB}}$ is indeed a unitary matrix that satisfies the constraint in (\ref{eq44}). Furthermore, we have
\begin{equation}\label{eq56}
  {\mathbf{F}}_{BB,2}^H{\mathbf{F}}_{RF}^H{{\mathbf{F}}_{BS}} = \sqrt {\frac{P}{{K{N_t}}}} {{{{\mathbf{\tilde V}}}_n^H}}{\mathbf{\tilde U}}_n^H{\mathbf{F}}_{RF}^H{{\mathbf{F}}_{BS}} = {\mathbf{0}},
\end{equation}
which suggests that
\begin{equation}\label{eq57}
  {\mathbf{F}}_{BS}^H{{\mathbf{F}}_{RF}}{{\mathbf{F}}_{BB,2}} = {\left( {{\mathbf{\tilde H}}{{{\mathbf{\tilde H}}}^H}} \right)^{ - 1}}{\mathbf{\tilde H}}{{\mathbf{F}}_{RF}}{{\mathbf{F}}_{BB,2}} = {\mathbf{0}}.
\end{equation}
By multiplying the above equation with ${{\mathbf{B}}\left( \Phi  \right)}{{\mathbf{\tilde H}}{{{\mathbf{\tilde H}}}^H}}$, we have
\begin{equation}\label{eq58}
{\mathbf{B}}\left( \Phi  \right){\mathbf{\tilde H}}{{\mathbf{F}}_{RF}}{{\mathbf{F}}_{BB,2}} = {\mathbf{H}}{{\mathbf{F}}_{RF}}{{\mathbf{F}}_{BB,2}} = {\mathbf{0}}.
\end{equation}
This completes the proof.


%




\ifCLASSOPTIONcaptionsoff
  \newpage
\fi


\bibliographystyle{IEEEtran}
\bibliography{IEEEabrv,RadCom}

\begin{thebibliography}{100}
\providecommand{\url}[1]{#1}
\csname url@samestyle\endcsname
\providecommand{\newblock}{\relax}
\providecommand{\bibinfo}[2]{#2}
\providecommand{\BIBentrySTDinterwordspacing}{\spaceskip=0pt\relax}
\providecommand{\BIBentryALTinterwordstretchfactor}{4}
\providecommand{\BIBentryALTinterwordspacing}{\spaceskip=\fontdimen2\font plus
\BIBentryALTinterwordstretchfactor\fontdimen3\font minus
  \fontdimen4\font\relax}
\providecommand{\BIBforeignlanguage}[2]{{%
\expandafter\ifx\csname l@#1\endcsname\relax
\typeout{** WARNING: IEEEtran.bst: No hyphenation pattern has been}%
\typeout{** loaded for the language `#1'. Using the pattern for}%
\typeout{** the default language instead.}%
\else
\language=\csname l@#1\endcsname
\fi
#2}}
\providecommand{\BIBdecl}{\relax}
\BIBdecl

\bibitem{UK_spectrum}
\BIBentryALTinterwordspacing
BBC. (2015) {P}rice hike for {UK} mobile spectrum. [Online]. Available:
  \url{http://www.bbc.co.uk/news/technology-34346822}
\BIBentrySTDinterwordspacing

\bibitem{GM_spectrum}
\BIBentryALTinterwordspacing
A.~Morris. (2015) German spectrum auction raises more than 5{B}. [Online].
  Available:
  \url{https://www.fiercewireless.com/europe/german-spectrum-auction-raises-more-than-eu5b}
\BIBentrySTDinterwordspacing

\bibitem{US_spectrum}
\BIBentryALTinterwordspacing
S.~Riaz. (2019) {US} completes first 5{G} auction. [Online]. Available:
  \url{https://www.mobileworldlive.com/featured-content/top-three/us-completes-first-5g-auction/}
\BIBentrySTDinterwordspacing

\bibitem{Connected_devices}
\BIBentryALTinterwordspacing
P.~Brown. (2016) 75.4 billion devices connected to the internet of things by
  2025. [Online]. Available:
  \url{https://electronics360.globalspec.com/article/6551/75-4-billion-devices-connected-to-the-internet-of-things-by-2025}
\BIBentrySTDinterwordspacing

\bibitem{radar_spectrum}
H.~Griffiths, L.~Cohen, S.~Watts, E.~Mokole, C.~Baker, M.~Wicks, and S.~Blunt,
  ``Radar spectrum engineering and management: Technical and regulatory
  issues,'' \emph{Proc. IEEE}, vol. 103, no.~1, pp. 85--102, 2015.

\bibitem{FCC_broad}
\BIBentryALTinterwordspacing
{FCC}. (2010) Connecting {A}merica: {T}he national broadband plan. [Online].
  Available: \url{https://www.fcc.gov/general/national-broadband-plan}
\BIBentrySTDinterwordspacing

\bibitem{NSF_specees}
\BIBentryALTinterwordspacing
{NSF}. (2016) Spectrum efficiency, energy efficiency, and security (spec{EES}):
  {E}nabling spectrum for all. [Online]. Available:
  \url{https://www.nsf.gov/pubs/2016/nsf16616/nsf16616.htm}
\BIBentrySTDinterwordspacing

\bibitem{SSPARC}
\BIBentryALTinterwordspacing
{DARPA}. (2016) Shared spectrum access for radar and communications ({SSPARC}).
  [Online]. Available:
  \url{https://www.darpa.mil/program/shared-spectrum-access-for-radar-and-communications}
\BIBentrySTDinterwordspacing

\bibitem{Ofcom_PSSR}
\BIBentryALTinterwordspacing
{{O}fcom}. (2015) Public sector spectrum release ({PSSR}): {A}ward of the 2.3
  {GH}z and 3.4 {GH}z bands. [Online]. Available:
  \url{https://www.ofcom.org.uk/consultations-and-statements/category-1/2.3-3.4-ghz-auction-design}
\BIBentrySTDinterwordspacing

\bibitem{UK_radar_plannning}
\BIBentryALTinterwordspacing
CAA. Public sector spectrum release programme: Radar planning and spectrum
  sharing in the 2.7-2.9{GH}z bands. [Online]. Available:
  \url{https://www.caa.co.uk/Commercial-industry/Airspace/Communication-navigation-and-surveillance/Spectrum/Public-sector-spectrum-release-programme/}
\BIBentrySTDinterwordspacing

\bibitem{7782415}
B.~{Paul}, A.~R. {Chiriyath}, and D.~W. {Bliss}, ``Survey of {RF}
  communications and sensing convergence research,'' \emph{IEEE Access},
  vol.~5, pp. 252--270, Dec 2017.

\bibitem{8246850}
H.~{Wymeersch}, G.~{Seco-Granados}, G.~{Destino}, D.~{Dardari}, and
  F.~{Tufvesson}, ``5{G} mmwave positioning for vehicular networks,''
  \emph{IEEE Wireless Commun.}, vol.~24, no.~6, pp. 80--86, Dec 2017.

\bibitem{7060497}
C.~{Yang} and H.~{Shao}, ``{WiFi}-based indoor positioning,'' \emph{IEEE
  Commun. Mag.}, vol.~53, no.~3, pp. 150--157, Mar 2015.

\bibitem{5545182}
S.~D. {Blunt}, P.~{Yatham}, and J.~{Stiles}, ``Intrapulse radar-embedded
  communications,'' \emph{IEEE Trans. Aerosp. Electron. Syst.}, vol.~46, no.~3,
  pp. 1185--1200, Jul 2010.

\bibitem{rad_lte}
H.~Wang, J.~T. Johnson, and C.~J. Baker, ``Spectrum sharing between
  communications and {ATC} radar systems,'' \emph{IET Radar Sonar Navig.},
  vol.~11, no.~6, pp. 994--1001, Jul 2017.

\bibitem{7462190}
J.~H. {Reed}, A.~W. {Clegg}, A.~V. {Padaki}, T.~{Yang}, R.~{Nealy},
  C.~{Dietrich}, C.~R. {Anderson}, and D.~M. {Mearns}, ``On the co-existence of
  {TD-LTE} and radar over 3.5 {GH}z band: {A}n experimental study,'' \emph{IEEE
  Wireless Commun. Lett.}, vol.~5, no.~4, pp. 368--371, Aug 2016.

\bibitem{rad_wifi}
F.~Hessar and S.~Roy, ``Spectrum sharing between a surveillance radar and
  secondary {W}i-{F}i networks,'' \emph{IEEE Trans. Aerosp. Electron. Syst.},
  vol.~52, no.~3, pp. 1434--1448, June 2016.

\bibitem{wiki_wlan}
\BIBentryALTinterwordspacing
W.~Contributors. (2019) List of {WLAN} channels - {W}ikipedia, the free
  encyclopedia. [Online]. Available:
  \url{https://en.wikipedia.org/wiki/{L}ist\textunderscore of \textunderscore
  {WLAN}\textunderscore channels}
\BIBentrySTDinterwordspacing

\bibitem{7786130}
J.~{Choi}, V.~{Va}, N.~{Gonzalez-Prelcic}, R.~{Daniels}, C.~R. {Bhat}, and
  R.~W. {Heath}, ``Millimeter-wave vehicular communication to support massive
  automotive sensing,'' \emph{IEEE Commun. Mag.}, vol.~54, no.~12, pp.
  160--167, Dec 2016.

\bibitem{6736750}
W.~{Roh}, J.~{Seol}, J.~{Park}, B.~{Lee}, J.~{Lee}, Y.~{Kim}, J.~{Cho},
  K.~{Cheun}, and F.~{Aryanfar}, ``Millimeter-wave beamforming as an enabling
  technology for 5{G} cellular communications: {T}heoretical feasibility and
  prototype results,'' \emph{IEEE Commun. Mag.}, vol.~52, no.~2, pp. 106--113,
  Feb 2014.

\bibitem{radar_lte_delay}
\BIBentryALTinterwordspacing
Mobile{E}urope. Airport radars could delay {LTE} take-off in south east
  england. [Online]. Available:
  \url{https://www.mobileeurope.co.uk/news-analysis/airports-not-on-lte-radar}
\BIBentrySTDinterwordspacing

\bibitem{5888501}
J.~B. Kenney, ``Dedicated short-range communications ({DSRC}) standards in the
  united states,'' \emph{Proc. IEEE}, vol.~99, no.~7, pp. 1162--1182, Jul 2011.

\bibitem{6515173}
T.~S. Rappaport, S.~Sun, R.~Mayzus, H.~Zhao, Y.~Azar, K.~Wang, G.~N. Wong,
  J.~K. Schulz, M.~Samimi, and F.~Gutierrez, ``Millimeter wave mobile
  communications for 5{G} cellular: {I}t will work!'' \emph{IEEE Access},
  vol.~1, pp. 335--349, May 2013.

\bibitem{7400949}
R.~W. {Heath}, N.~{González-Prelcic}, S.~{Rangan}, W.~{Roh}, and A.~M.
  {Sayeed}, ``An overview of signal processing techniques for millimeter wave
  {MIMO} systems,'' \emph{IEEE J. Sel. Topics Signal Process.}, vol.~10, no.~3,
  pp. 436--453, Apr 2016.

\bibitem{7303962}
C.~{Xu}, B.~{Firner}, Y.~{Zhang}, and R.~E. {Howard}, ``The case for efficient
  and robust {RF}-based device-free localization,'' \emph{{IEEE} Trans. Mobile
  Comput.}, vol.~15, no.~9, pp. 2362--2375, Sep 2016.

\bibitem{6042868}
C.~{Feng}, W.~S.~A. {Au}, S.~{Valaee}, and Z.~{Tan},
  ``Received-signal-strength-based indoor positioning using compressive
  sensing,'' \emph{IEEE Trans. Mobile Comput.}, vol.~11, no.~12, pp.
  1983--1993, Dec 2012.

\bibitem{6244790}
K.~{Wu}, J.~{Xiao}, Y.~{Yi}, D.~{Chen}, X.~{Luo}, and L.~M. {Ni}, ``{CSI}-based
  indoor localization,'' \emph{IEEE Trans. Parallel Distrib. Syst.}, vol.~24,
  no.~7, pp. 1300--1309, Jul 2013.

\bibitem{8360863}
B.~{Tan}, Q.~{Chen}, K.~{Chetty}, K.~{Woodbridge}, W.~{Li}, and R.~{Piechocki},
  ``Exploiting {W}i{F}i channel state information for residential healthcare
  informatics,'' \emph{IEEE Commun. Mag.}, vol.~56, no.~5, pp. 130--137, May
  2018.

\bibitem{7944276}
F.~{Fioranelli}, M.~{Ritchie}, and H.~{Griffiths}, ``Bistatic human
  micro-{D}oppler signatures for classification of indoor activities,'' in
  \emph{2017 IEEE Radar Conference (RadarConf)}, May 2017, pp. 0610--0615.

\bibitem{7426551}
M.~G. {Amin}, Y.~D. {Zhang}, F.~{Ahmad}, and K.~C.~D. {Ho}, ``Radar signal
  processing for elderly fall detection: {T}he future for in-home monitoring,''
  \emph{IEEE Signal Process. Mag.}, vol.~33, no.~2, pp. 71--80, Mar 2016.

\bibitem{7046290}
Q.~{Wu}, Y.~D. {Zhang}, W.~{Tao}, and M.~G. {Amin}, ``Radar-based fall
  detection based on {D}oppler time–frequency signatures for assisted
  living,'' \emph{IET Radar Sonar Navig.}, vol.~9, no.~2, pp. 164--172, 2015.

\bibitem{google_soli}
\BIBentryALTinterwordspacing
C.~{Dubois}. (2019) Google {ATAP} moves forward with radar touch tech with
  {FCC} waiver. [Online]. Available:
  \url{https://www.allaboutcircuits.com/news/Google-ATAP-Project-Soli-radar-touch-sensor-technology-FCC-waiver/}
\BIBentrySTDinterwordspacing

\bibitem{8531711}
S.~{Zhang}, Y.~{Zeng}, and R.~{Zhang}, ``Cellular-enabled uav communication: A
  connectivity-constrained trajectory optimization perspective,'' \emph{IEEE
  Trans. Commun.}, vol.~67, no.~3, pp. 2580--2604, Mar 2019.

\bibitem{1428700}
A.~{Ryan}, M.~{Zennaro}, A.~{Howell}, R.~{Sengupta}, and J.~K. {Hedrick}, ``An
  overview of emerging results in cooperative {UAV} control,'' in \emph{2004
  43rd IEEE Conference on Decision and Control (CDC)}, vol.~1, Dec 2004, pp.
  602--607.

\bibitem{7470933}
Y.~{Zeng}, R.~{Zhang}, and T.~J. {Lim}, ``Wireless communications with unmanned
  aerial vehicles: {O}pportunities and challenges,'' \emph{IEEE Commun. Mag.},
  vol.~54, no.~5, pp. 36--42, May 2016.

\bibitem{6646211}
N.~{Decarli}, F.~{Guidi}, and D.~{Dardari}, ``A novel joint {RFID} and radar
  sensor network for passive localization: Design and performance bounds,''
  \emph{IEEE J. Sel. Topics Signal Process.}, vol.~8, no.~1, pp. 80--95, Feb
  2014.

\bibitem{bio_sen}
G.~Fortino, M.~Pathan, and G.~D. Fatta, ``Body{C}loud: {I}ntegration of cloud
  computing and body sensor networks,'' in \emph{4th IEEE International
  Conference on Cloud Computing Technology and Science Proceedings}, 2012, pp.
  851--856.

\bibitem{6875553}
D.~W. {Bliss}, ``Cooperative radar and communications signaling: {T}he
  estimation and information theory odd couple,'' in \emph{2014 IEEE Radar
  Conference}, May 2014, pp. 0050--0055.

\bibitem{858893}
P.~K. {Hughes} and J.~Y. {Choe}, ``Overview of advanced multifunction {RF}
  system ({AMRFS}),'' in \emph{Proceedings 2000 IEEE International Conference
  on Phased Array Systems and Technology (Cat. No.00TH8510)}, May 2000, pp.
  21--24.

\bibitem{1406306}
G.~C. {Tavik}, C.~L. {Hilterbrick}, J.~B. {Evins}, J.~J. {Alter}, J.~G.
  {Crnkovich}, J.~W. {de Graaf}, W.~{Habicht}, G.~P. {Hrin}, S.~A. {Lessin},
  D.~C. {Wu}, and S.~M. {Hagewood}, ``The advanced multifunction {RF}
  concept,'' \emph{IEEE Trans. Microw. Theory Technol.}, vol.~53, no.~3, pp.
  1009--1020, Mar 2005.

\bibitem{6127573}
J.~A. {Molnar}, I.~{Corretjer}, and G.~{Tavik}, ``Integrated topside -
  integration of narrowband and wideband array antennas for shipboard
  communications,'' in \emph{2011 - MILCOM 2011 Military Communications
  Conference}, Nov 2011, pp. 1802--1807.

\bibitem{1677946}
R.~W. {Beard}, T.~W. {McLain}, D.~B. {Nelson}, D.~{Kingston}, and
  D.~{Johanson}, ``Decentralized cooperative aerial surveillance using
  fixed-wing miniature {UAV}s,'' \emph{Proc. IEEE}, vol.~94, no.~7, pp.
  1306--1324, Jul 2006.

\bibitem{6105461}
R.~{Schneiderman}, ``Unmanned drones are flying high in the military/aerospace
  sector [special reports],'' \emph{IEEE Signal Process. Mag.}, vol.~29, no.~1,
  pp. 8--11, Jan 2012.

\bibitem{7935430}
Z.~R. {Bogdanowicz}, ``Flying swarm of drones over circulant digraph,''
  \emph{IEEE Trans. Aerosp. Electron. Syst.}, vol.~53, no.~6, pp. 2662--2670,
  Dec 2017.

\bibitem{8490190}
S.~{Winkler}, S.~{Zeadally}, and K.~{Evans}, ``Privacy and civilian drone use:
  {T}he need for further regulation,'' \emph{IEEE Security Privacy}, vol.~16,
  no.~5, pp. 72--80, Sep 2018.

\bibitem{5592987}
D.~B. {Ramos}, D.~S. {Loubach}, and A.~M. {da Cunha}, ``Developing a
  distributed real-time monitoring system to track {UAV}s,'' \emph{IEEE Trans.
  Aerosp. Electron. Syst.}, vol.~25, no.~9, pp. 18--25, Sep 2010.

\bibitem{8624565}
S.~{Zhang}, H.~{Zhang}, B.~{Di}, and L.~{Song}, ``Cellular {UAV}-to-{X}
  communications: {D}esign and optimization for multi-{UAV} networks,''
  \emph{IEEE Trans. Wireless Commun.}, vol.~18, no.~2, pp. 1346--1359, Feb
  2019.

\bibitem{1146255}
A.~{Polydoros} and K.~{Woo}, ``{LPI} detection of frequency-hopping signals
  using autocorrelation techniques,'' \emph{IEEE J. Sel. Areas Commun.},
  vol.~3, no.~5, pp. 714--726, Sep 1985.

\bibitem{1146256}
A.~{Polydoros} and C.~{Weber}, ``Detection performance considerations for
  direct-sequence and time-hopping {LPI} waveforms,'' \emph{IEEE J. Sel. Areas
  Commun.}, vol.~3, no.~5, pp. 727--744, Sep 1985.

\bibitem{6081358}
S.~D. {Blunt}, J.~G. {Metcalf}, C.~R. {Biggs}, and E.~{Perrins}, ``Performance
  characteristics and metrics for intra-pulse radar-embedded communication,''
  \emph{IEEE J. Sel. Areas Commun}, vol.~29, no.~10, pp. 2057--2066, Dec 2011.

\bibitem{7376230}
D.~{Ciuonzo}, A.~{De Maio}, G.~{Foglia}, and M.~{Piezzo}, ``Intrapulse
  radar-embedded communications via multiobjective optimization,'' \emph{IEEE
  Trans. Aerosp. Electron. Syst.}, vol.~51, no.~4, pp. 2960--2974, Oct 2015.

\bibitem{7944264}
S.~{Brisken}, M.~{Moscadelli}, V.~{Seidel}, and C.~{Schwark}, ``Passive radar
  imaging using {DVB-S2},'' in \emph{2017 IEEE Radar Conference (RadarConf)},
  May 2017, pp. 0552--0556.

\bibitem{passive_radar_book}
H.~Griffiths and C.~J. Baker, \emph{An Introduction to Passive Radar}.\hskip
  1em plus 0.5em minus 0.4em\relax Artech House, 2017.

\bibitem{7962141}
B.~K. {Chalise}, M.~G. {Amin}, and B.~{Himed}, ``Performance tradeoff in a
  unified passive radar and communications system,'' \emph{IEEE Signal Process.
  Lett.}, vol.~24, no.~9, pp. 1275--1279, Sep 2017.

\bibitem{4557030}
L.~S. {Wang}, J.~P. {Mcgeehan}, C.~{Williams}, and A.~{Doufexi}, ``Application
  of cooperative sensing in radar-communications coexistence,'' \emph{IET
  Commun.}, vol.~2, no.~6, pp. 856--868, Jul 2008.

\bibitem{6331681}
R.~{Saruthirathanaworakun}, J.~M. {Peha}, and L.~M. {Correia}, ``Opportunistic
  sharing between rotating radar and cellular,'' \emph{IEEE J. Sel. Areas
  Commun.}, vol.~30, no.~10, pp. 1900--1910, Nov 2012.

\bibitem{4350230}
J.~{Li} and P.~{Stoica}, ``{MIMO} radar with colocated antennas,'' \emph{IEEE
  Signal Process. Mag.}, vol.~24, no.~5, pp. 106--114, Sep 2007.

\bibitem{li2008mimo}
J.~Li and P.~Stoica, \emph{{MIMO} radar signal processing}.\hskip 1em plus
  0.5em minus 0.4em\relax John Wiley \& Sons, 2008.

\bibitem{7814210}
J.~A. {Mahal}, A.~{Khawar}, A.~{Abdelhadi}, and T.~C. {Clancy}, ``Spectral
  coexistence of {MIMO} radar and {MIMO} cellular system,'' \emph{IEEE Trans.
  Aerosp. Electron. Syst.}, vol.~53, no.~2, pp. 655--668, Apr 2017.

\bibitem{7953658}
B.~{Li} and A.~P. {Petropulu}, ``Joint transmit designs for coexistence of
  {MIMO} wireless communications and sparse sensing radars in clutter,''
  \emph{IEEE Trans. Aerosp. Electron. Syst.}, vol.~53, no.~6, pp. 2846--2864,
  Dec 2017.

\bibitem{ICSI_est}
\BIBentryALTinterwordspacing
F.~Liu, A.~Garcia-Rodriguez, C.~Masouros, and G.~Geraci. (2018) Interfering
  channel estimation in radar-cellular coexistence: {H}ow much information do
  we need? [Online]. Available: \url{https://arxiv.org/abs/1807.04000}
\BIBentrySTDinterwordspacing

\bibitem{7089157}
A.~{Khawar}, A.~{Abdelhadi}, and C.~{Clancy}, ``Target detection performance of
  spectrum sharing {MIMO} radars,'' \emph{IEEE Sensors J.}, vol.~15, no.~9, pp.
  4928--4940, Sep 2015.

\bibitem{6831613}
A.~{Babaei}, W.~H. {Tranter}, and T.~{Bose}, ``A nullspace-based precoder with
  subspace expansion for radar/communications coexistence,'' in \emph{2013 IEEE
  Global Communications Conference (GLOBECOM)}, Dec 2013, pp. 3487--3492.

\bibitem{7470514}
B.~{Li}, A.~P. {Petropulu}, and W.~{Trappe}, ``Optimum co-design for spectrum
  sharing between matrix completion based {MIMO} radars and a {MIMO}
  communication system,'' \emph{IEEE Trans. Signal Process.}, vol.~64, no.~17,
  pp. 4562--4575, Sep 2016.

\bibitem{8048004}
L.~{Zheng}, M.~{Lops}, X.~{Wang}, and E.~{Grossi}, ``Joint design of overlaid
  communication systems and pulsed radars,'' \emph{IEEE Trans. Signal
  Process.}, vol.~66, no.~1, pp. 139--154, Jan 2018.

\bibitem{7898445}
F.~{Liu}, C.~{Masouros}, A.~{Li}, and T.~{Ratnarajah}, ``Robust {MIMO}
  beamforming for cellular and radar coexistence,'' \emph{IEEE Wireless Commun.
  Lett.}, vol.~6, no.~3, pp. 374--377, Jun 2017.

\bibitem{8445973}
Y.~{Cui}, V.~{Koivunen}, and X.~{Jing}, ``Interference alignment based spectrum
  sharing for {MIMO} radar and communication systems,'' in \emph{2018 IEEE 19th
  International Workshop on Signal Processing Advances in Wireless
  Communications (SPAWC)}, Jun 2018, pp. 1--5.

\bibitem{8355705}
F.~{Liu}, C.~{Masouros}, A.~{Li}, T.~{Ratnarajah}, and J.~{Zhou}, ``{MIMO}
  radar and cellular coexistence: A power-efficient approach enabled by
  interference exploitation,'' \emph{IEEE Trans. Signal Process.}, vol.~66,
  no.~14, pp. 3681--3695, Jul 2018.

\bibitem{7103338}
C.~{Masouros} and G.~{Zheng}, ``Exploiting known interference as green signal
  power for downlink beamforming optimization,'' \emph{IEEE Trans. Signal
  Process.}, vol.~63, no.~14, pp. 3628--3640, Jul 2015.

\bibitem{8233171}
L.~{Zheng}, M.~{Lops}, and X.~{Wang}, ``Adaptive interference removal for
  uncoordinated radar/communication coexistence,'' \emph{IEEE J. Sel. Topics
  Signal Process.}, vol.~12, no.~1, pp. 45--60, Feb 2018.

\bibitem{8332962}
N.~{Nartasilpa}, A.~{Salim}, D.~{Tuninetti}, and N.~{Devroye}, ``Communications
  system performance and design in the presence of radar interference,''
  \emph{IEEE Trans. Commun.}, vol.~66, no.~9, pp. 4170--4185, Sep 2018.

\bibitem{7279172}
A.~R. {Chiriyath}, B.~{Paul}, G.~M. {Jacyna}, and D.~W. {Bliss}, ``Inner bounds
  on performance of radar and communications co-existence,'' \emph{IEEE Trans.
  Signal Process.}, vol.~64, no.~2, pp. 464--474, Jan 2016.

\bibitem{7131098}
J.~R. {Guerci}, R.~M. {Guerci}, A.~{Lackpour}, and D.~{Moskowitz}, ``Joint
  design and operation of shared spectrum access for radar and
  communications,'' in \emph{2015 IEEE Radar Conference (RadarCon)}, May 2015,
  pp. 0761--0766.

\bibitem{kay1998fundamentals}
S.~M. Kay, \emph{Fundamentals of Statistical Signal Processing, {V}ol. {I}:
  {E}stimation Theory}.\hskip 1em plus 0.5em minus 0.4em\relax Englewood
  Cliffs, NJ, USA: Prentice Hall, 1998.

\bibitem{chiriyath2017radar}
A.~R. Chiriyath, B.~Paul, and D.~W. Bliss, ``Radar-communications convergence:
  {C}oexistence, cooperation, and co-design,'' \emph{IEEE Trans. Cogn. Commun.
  Netw.}, vol.~3, no.~1, pp. 1--12, 2017.

\bibitem{8448783}
Y.~{Rong}, A.~R. {Chiriyath}, and D.~W. {Bliss}, ``{MIMO} radar and
  communications spectrum sharing: A multiple-access perspective,'' in
  \emph{2018 IEEE 10th Sensor Array and Multichannel Signal Processing Workshop
  (SAM)}, Jul 2018, pp. 272--276.

\bibitem{mealey1963method}
R.~M. Mealey, ``A method for calculating error probabilities in a radar
  communication system,'' \emph{IEEE Trans. Space Electron. Telemetry}, vol.~9,
  no.~2, pp. 37--42, Jun 1963.

\bibitem{roberton2003integrated}
M.~Roberton and E.~R. Brown, ``Integrated radar and communications based on
  chirped spread-spectrum techniques,'' in \emph{Microwave Symposium Digest,
  2003 IEEE MTT-S International}, vol.~1, 2003, pp. 611--614.

\bibitem{saddik2007ultra}
G.~N. Saddik, R.~S. Singh, and E.~R. Brown, ``Ultra-wideband multifunctional
  communications/radar system,'' \emph{IEEE Trans. Microw. Theory Technol.},
  vol.~55, no.~7, pp. 1431--1437, Jul 2007.

\bibitem{jamil2008integrated}
M.~Jamil, H.-J. Zepernick, and M.~I. Pettersson, ``On integrated radar and
  communication systems using {O}ppermann sequences,'' in \emph{Proc. IEEE
  Military Commun.}, 2008, pp. 1--6.

\bibitem{han2013joint}
L.~Han and K.~Wu, ``Joint wireless communication and radar sensing
  systems-state of the art and future prospects,'' \emph{IET Microw. Antennas
  Propag.}, vol.~7, no.~11, pp. 876--885, 2013.

\bibitem{garmatyuk2011multifunctional}
D.~Garmatyuk, J.~Schuerger, and K.~Kauffman, ``Multifunctional software-defined
  radar sensor and data communication system,'' \emph{IEEE Sensors J.},
  vol.~11, no.~1, pp. 99--106, Jan 2011.

\bibitem{sturm2011waveform}
C.~Sturm and W.~Wiesbeck, ``Waveform design and signal processing aspects for
  fusion of wireless communications and radar sensing,'' \emph{Proc. IEEE},
  vol.~99, no.~7, pp. 1236--1259, Jul 2011.

\bibitem{7485314}
D.~{Gaglione}, C.~{Clemente}, C.~V. {Ilioudis}, A.~R. {Persico}, I.~K.
  {Proudler}, and J.~J. {Soraghan}, ``Fractional fourier based waveform for a
  joint radar-communication system,'' in \emph{2016 IEEE Radar Conference
  (RadarConf)}, May 2016, pp. 1--6.

\bibitem{330368}
L.~B. {Almeida}, ``The fractional {F}ourier transform and time-frequency
  representations,'' \emph{IEEE Trans. Signal Process.}, vol.~42, no.~11, pp.
  3084--3091, Nov 1994.

\bibitem{7347464}
A.~{Hassanien}, M.~G. {Amin}, Y.~D. {Zhang}, and F.~{Ahmad}, ``Dual-function
  radar-communications: Information embedding using sidelobe control and
  waveform diversity,'' \emph{IEEE Trans. Signal Process.}, vol.~64, no.~8, pp.
  2168--2181, Apr 2016.

\bibitem{7485066}
A.~{Hassanien}, M.~G. {Amin}, Y.~D. {Zhang}, F.~{Ahmad}, and B.~{Himed},
  ``Non-coherent psk-based dual-function radar-communication systems,'' in
  \emph{2016 IEEE Radar Conference (RadarConf)}, May 2016, pp. 1--6.

\bibitem{7485316}
E.~{BouDaher}, A.~{Hassanien}, E.~{Aboutanios}, and M.~G. {Amin}, ``Towards a
  dual-function mimo radar-communication system,'' in \emph{2016 IEEE Radar
  Conference (RadarConf)}, May 2016, pp. 1--6.

\bibitem{8288677}
F.~{Liu}, C.~{Masouros}, A.~{Li}, H.~{Sun}, and L.~{Hanzo}, ``{MU-MIMO}
  communications with {MIMO} radar: {F}rom co-existence to joint
  transmission,'' \emph{IEEE Trans. Wireless Commun.}, vol.~17, no.~4, pp.
  2755--2770, Apr 2018.

\bibitem{8386661}
F.~{Liu}, L.~{Zhou}, C.~{Masouros}, A.~{Li}, W.~{Luo}, and A.~{Petropulu},
  ``Toward dual-functional radar-communication systems: {O}ptimal waveform
  design,'' \emph{IEEE Trans. Signal Process.}, vol.~66, no.~16, pp.
  4264--4279, Aug 2018.

\bibitem{sspd_2019}
F.~{Liu}, C.~{Masouros}, and H.~{Griffiths}, ``Dual-functional
  radar-communication waveform design under constant-modulus and orthogonality
  constraints,'' in \emph{2019 Sensor Signal Processing for Defence (SSPD)},
  May 2019.

\bibitem{8057284}
R.~C. {Daniels}, E.~R. {Yeh}, and R.~W. {Heath}, ``Forward collision vehicular
  radar with {IEEE} 802.11: Feasibility demonstration through measurements,''
  \emph{IEEE Trans. Veh. Technol.}, vol.~67, no.~2, pp. 1404--1416, Feb 2018.

\bibitem{8114253}
P.~{Kumari}, J.~{Choi}, N.~{González-Prelcic}, and R.~W. {Heath}, ``{IEEE}
  802.11ad-based radar: {A}n approach to joint vehicular communication-radar
  system,'' \emph{IEEE Trans. Veh. Technol.}, vol.~67, no.~4, pp. 3012--3027,
  Apr 2018.

\bibitem{8309274}
E.~{Grossi}, M.~{Lops}, L.~{Venturino}, and A.~{Zappone}, ``Opportunistic radar
  in {IEEE} 802.11ad networks,'' \emph{IEEE Trans. Signal Process.}, vol.~66,
  no.~9, pp. 2441--2454, May 2018.

\bibitem{7010533}
S.~{Han}, C.~{I}, Z.~{Xu}, and C.~{Rowell}, ``Large-scale antenna systems with
  hybrid analog and digital beamforming for millimeter wave 5{G},'' \emph{IEEE
  Commun. Mag.}, vol.~53, no.~1, pp. 186--194, Jan 2015.

\bibitem{8030501}
A.~F. Molisch, V.~V. Ratnam, S.~Han, Z.~Li, S.~L.~H. Nguyen, L.~Li, and
  K.~Haneda, ``Hybrid beamforming for massive {MIMO}: {A} survey,'' \emph{IEEE
  Commun. Mag.}, vol.~55, no.~9, pp. 134--141, Sep 2017.

\bibitem{8550811}
J.~A. {Zhang}, X.~{Huang}, Y.~J. {Guo}, J.~{Yuan}, and R.~W. {Heath},
  ``Multibeam for joint communication and radar sensing using steerable analog
  antenna arrays,'' \emph{IEEE Trans. Veh. Technol.}, vol.~68, no.~1, pp.
  671--685, Jan 2019.

\bibitem{4655353}
L.~{Xu}, J.~{Li}, and P.~{Stoica}, ``Target detection and parameter estimation
  for {MIMO} radar systems,'' \emph{IEEE Trans. Aerosp. Electron. Syst.},
  vol.~44, no.~3, pp. 927--939, Jul 2008.

\bibitem{5419124}
A.~{Hassanien} and S.~A. {Vorobyov}, ``Phased-{MIMO} radar: A tradeoff between
  phased-array and mimo radars,'' \emph{IEEE Trans. Signal Process.}, vol.~58,
  no.~6, pp. 3137--3151, Jun 2010.

\bibitem{1316398}
E.~{Fishler}, A.~{Haimovich}, R.~{Blum}, D.~{Chizhik}, L.~{Cimini}, and
  R.~{Valenzuela}, ``{MIMO} radar: an idea whose time has come,'' in
  \emph{Proceedings of the 2004 IEEE Radar Conference (IEEE Cat.
  No.04CH37509)}, Apr 2004, pp. 71--78.

\bibitem{5109947}
F.~{Daum} and J.~{Huang}, ``{MIMO} radar: {S}nake oil or good idea?''
  \emph{IEEE Trans. Aerosp. Electron. Syst.}, vol.~24, no.~5, pp. 8--12, May
  2009.

\bibitem{6104178}
D.~{Wilcox} and M.~{Sellathurai}, ``On {MIMO} radar subarrayed transmit
  beamforming,'' \emph{IEEE Trans. Signal Process.}, vol.~60, no.~4, pp.
  2076--2081, Apr 2012.

\bibitem{4276989}
P.~Stoica, J.~Li, and Y.~Xie, ``On probing signal design for {MIMO} radar,''
  \emph{IEEE Trans. Signal Process.}, vol.~55, no.~8, pp. 4151--4161, Aug 2007.

\bibitem{8052157}
D.~Zhang, Y.~Wang, X.~Li, and W.~Xiang, ``Hybridly connected structure for
  hybrid beamforming in mmwave massive {MIMO} systems,'' \emph{IEEE Trans.
  Commun.}, vol.~66, no.~2, pp. 662--674, Feb 2018.

\bibitem{7888145}
N.~González-Prelcic, R.~Méndez-Rial, and R.~W. Heath, ``Radar aided beam
  alignment in mm{W}ave {V}2{I} communications supporting antenna diversity,''
  in \emph{Proc. Information Theory and Applications Workshop (ITA)}, Jan 2016,
  pp. 1--7.

\bibitem{6994289}
H.~{Jiang}, J.~{Zhang}, and K.~M. {Wong}, ``Joint {DOD} and {DOA} estimation
  for bistatic {MIMO} radar in unknown correlated noise,'' \emph{IEEE Trans.
  Veh. Technol.}, vol.~64, no.~11, pp. 5113--5125, Nov 2015.

\bibitem{7504377}
G.~{Berardinelli}, K.~I. {Pedersen}, F.~{Frederiksen}, and P.~{Mogensen}, ``On
  the guard period design in 5{G} {TDD} wide area,'' in \emph{2016 IEEE 83rd
  Vehicular Technology Conference (VTC Spring)}, May 2016, pp. 1--5.

\bibitem{tse2005fundamentals}
D.~Tse and P.~Viswanath, \emph{Fundamentals of wireless communication}.\hskip
  1em plus 0.5em minus 0.4em\relax Cambridge university press, 2005.

\bibitem{7450660}
O.~Aldayel, V.~Monga, and M.~Rangaswamy, ``Successive {QCQP} refinement for
  {MIMO} radar waveform design under practical constraints,'' \emph{IEEE Trans.
  Signal Process.}, vol.~64, no.~14, pp. 3760--3774, Jul 2016.

\bibitem{1143830}
R.~{Schmidt}, ``Multiple emitter location and signal parameter estimation,''
  \emph{IEEE Trans. Antennas Propag.}, vol.~34, no.~3, pp. 276--280, Mar 1986.

\bibitem{506612}
J.~{Li} and P.~{Stoica}, ``An adaptive filtering approach to spectral
  estimation and {SAR} imaging,'' \emph{IEEE Trans. Signal Process.}, vol.~44,
  no.~6, pp. 1469--1484, Jun 1996.

\bibitem{viklands2008algorithms}
T.~Viklands, ``Algorithms for the weighted orthogonal {P}rocrustes problem and
  other least squares problems,'' Ph.D. dissertation, Comput. Sci. Dept., Umea
  Univ., Umea, Sweden, 2008.

\bibitem{8327462}
A.~{Deligiannis}, A.~{Daniyan}, S.~{Lambotharan}, and J.~A. {Chambers},
  ``Secrecy rate optimizations for {MIMO} communication radar,'' \emph{IEEE
  Trans. Aerosp. Electron. Syst.}, vol.~54, no.~5, pp. 2481--2492, Oct 2018.

\bibitem{CHALISE2018282}
B.~K. Chalise and M.~G. Amin, ``Performance tradeoff in a unified system of
  communications and passive radar: {A} secrecy capacity approach,''
  \emph{Digital Signal Process.}, vol.~82, pp. 282 -- 293, 2018.

\bibitem{radar_privacy}
A.~{Dimas}, M.~A. {Clark}, K.~{Psounis}, and A.~P. {Petropulu}, ``On radar
  privacy in shared spectrum scenarios,'' in \emph{2019 IEEE International
  Conference on Acoustics, Speech and Signal Processing (ICASSP)}, May 2019.

\bibitem{8642926}
S.~H. {Dokhanchi}, B.~S. {Mysore}, K.~V. {Mishra}, and B.~{Ottersten}, ``A
  mm{W}ave automotive joint radar-communications system,'' \emph{IEEE Trans.
  Aerosp. Electron. Syst.}, pp. 1--1, 2019.

\bibitem{8070342}
S.~{Shahi}, D.~{Tuninetti}, and N.~{Devroye}, ``On the capacity of the {AWGN}
  channel with additive radar interference,'' \emph{IEEE Trans. Commun.},
  vol.~66, no.~2, pp. 629--643, Feb 2018.

\end{thebibliography}
\end{document}